# Antihydrogen, probed with classical polarization dependent wavelength (PDW) shifts in the Lyman series, proves QFT inconsistent on antimatter

G. Van Hooydonk, Ghent University, Faculty of Sciences, Krijgslaan 281, B-9000 Ghent, Belgium

**Abstract**. Bound state QED uses the Sommerfeld-Dirac double square root equation to obtain quartics (Mexican hat or double well curves), which can give away left-right symmetry or chiral behavior for particle systems as in the SM. We now show that errors of Bohr 2D fermion theory are classical H polarization dependent wavelength (PDW) shifts. The observed H line spectrum exhibits a quartic with critical n-values for phase ½π (90°) and π (180°): phase ½π refers to circular H polarization (chiral behavior), phase π to linear H polarization and inversion on the Coulomb field axis. These signatures probe H polarization with 2 natural, mutually exclusive hydrogen quantum states ±1, i.e. H and H̲. The H̲ signatures are consistent with polarization angles or phases, hidden in de Broglie's standing wave equation, which derives from Compton's early experiments on sinusoidal wavelength shifts. Positive pressures in the matter-well (H) become negative in the antimatter-well (H̲), where they are linked with dark matter. We refute the ban on natural H̲ and prove why QED, a quartic generating quantum field theory, classifies as inconsistent on neutral antimatter.
    arXiv **physics/0612141**         Pacs:

## Introduction

The widely accepted ban on natural H̲, proclaimed by theorists, led experimentalists to measure interval 1S-2S for artificially produced H̲. Yet, we found unambiguous spectral signatures for H̲ in available H line and $H_2$ band spectra [1-5]. This H̲-controversy could be settled once and for all with an ab initio H polarization theory but no such theory exists. This evident failure of theory can never mean that H̲ does not exist or that a ban on H̲ is legitimate. On the contrary, since H̲ is field-inverted H by definition, polarization dependent wavelength (PDW) shifts can interfere. We prove that errors of Bohr H theory are classical sinusoidal polarization dependent (PDW) shifts, leading to spectral signatures for H̲ [1-3]. A Bohr-type H polarization theory accounts for PDW shifts (H̲-signatures) and leads to H boson behavior. With its ban on H̲, QFT proves inconsistent on neutral antimatter, which may have implications for dark matter too. This mistake on H̲ finds its origin in wrong interpretations of (i) work by Sommerfeld-Kratzer and by Compton-de Broglie, all published around 1920, and (ii) of the Lamb-shift, exposed in 1947.

## Generic sinusoidal appearance of polarization

Polarization of light, known for centuries, is a natural phenomenon, understood with a variable phase θ between 2 sinusoidal fields $a_0\cos\varphi + b_0\cos(\varphi+\theta)$. A 3D Poincaré (-Bloch) sphere polarization model uses poles ±1 and an equator to account for left-right or chiral properties of radiation: Cartesian frames for poles at +x and –x are the mutually exclusive pair (left- and right-handed) +x,+y,+z and –x,+y,+z. Any H theory based upon a constant phase like Bohr's orthogonal phase of 90° must always fail on H polarization, which requires fields out of phase instead of fields in phase.

Polarization of matter and its chirality prove difficult, although observing polarization must be rather straightforward with its sinusoidal dependence on phase θ, a fractional angle absolutely confined to domain $0 \leq \theta \leq 2\pi$. Properly scaled sinusoidal effects are easily retraceable, whatever their magnitude. A dimension-less cosine law for a numerical field f consisting of 2 sub-fields

$$f(\theta)=\pm\sqrt{[1+(a/b)^2 - 2(a/b)\cos\theta]} \approx \pm[1-(a/b)\cos\theta\ldots] \qquad (1a)$$

(square bracket version valid for b>>a), a convenient basis to get at polarization in a Poincaré sphere, implies oscillations or fluctuations of the metric. The square root in (1a) gives a tilted polarization ellipse and returns a circle for θ=90° or ½π and a=b. With b>>a like in X-ray experiments for non-resonant interactions, (1a) gives sinusoidal fluctuations in function of phase θ, as observed by Compton[1] a long time ago [6]. Writing (1a) as $[1-f(\theta)]/(a/b) \approx \cos\theta$ gives a cosine with generic asymptotes ±1. For the metric of the electron-proton Coulomb bond in composite H, numerical field (1a) gives two variants

$$f(\theta)=\pm 1/\sqrt{[1+(r_p/r_e)^2 - 2(r_p/r_e)\cos\theta]} \approx \pm[1+(r_p/r_e)\cos\theta\ldots] \qquad (1b)$$

$$f(\theta)=\pm 1/\sqrt{[1+(m_e/m_p)^2 - 2(m_e/m_p)\cos\theta]} \approx \pm[1+(m_e/m_p)\cos\theta\ldots] \qquad (1c)$$

Strictly spoken, (1b) is a pair of conjugated unit charges short and (1c) is deceptive on mass as a scalar[2]. Both variants are valid with internal field equations $m_e r_e = m_p r_p$ or $m_e/m_p = r_p/r_e$, the classical equilibrium conditions for bound states. Despite their gravitational basis, these are needed to fix the origin of the reference frame for composite H, without which the basic equations for H and its complementary sub-particles cannot even be formulated properly.

Polarization problems are not solved with STR and its orthogonal fields $f(90°)=\pm\sqrt{[1+(a/b)^2]}$. Although (1a) belongs to the classical 19th century Stokes-Maxwell-Poincaré view on polarization[3], bound state QED uses a Dirac-field approach, inspired by Einstein's STR equation

$$f(\alpha/n)=1/\sqrt{[1+(\alpha/n)^2]} = 1-\tfrac{1}{2}(\alpha/n)^2 + \tfrac{3}{8}(\alpha/n)^2 - \ldots \qquad (1d)$$

incompatible with (1a) needed for polarization, due to a lack of sinusoidal effects on the metric. How STR field (1d) had to be modified for H bound state QED is shown below. For full resonance of hν (photons, boson symmetry and polarization) and H energies $E_1$-$E_2$ (2 fermions) to be possible with

$$h\nu = E_1 - E_2 \qquad (1e)$$

atom H must be credited with boson symmetry. Since Bohr's H is of fermion type, solutions based upon (1a)-(1c) for H, involving oscillations typical for boson behavior, must be envisaged. Compton's results led to de Broglie's the standing wave equation as a condition for resonance (1e) to occur

$$2\pi r = k\lambda \qquad (1f)$$

which, in turn, led to non-relativistic Schrödinger wave mechanics and relativistic QFTs. In terms of field effects (1a-d), resonance condition (1f) hides a numerical quantized field ratio

$$r/\lambda = k/2\pi \qquad (1g)$$

---

[1] which makes Compton polarimetry important for elementary particle physics [7].
[2] Reduced mass using complementary sub-masses $m_H = m_e + m_p = +m_e + (m_H - m_e)$ is not a scalar (see below).
[3] later on complemented with Jones [8] and Mueller matrices [9]

suggesting that de Broglie's quantum number k is competitive with phase θ with maximum 2π. For polarization, this has drastic consequences, since rewriting de Broglie's relation (1g) as

$$r/(½\lambda) = k/\pi \text{ and } r/(¼\lambda) = k/(½\pi) \quad (1h)$$

gives a half wavelength plate with k equal to 180° or π, known to transform left- in right-handed radiation, while a quarter wavelength plate acts accordingly in terms of classical optics. Detecting critical H-states for a quantum number equal to ½π and ½π is therefore essential for H polarization.

**Bohr-type H polarization**

Without specifying sub-fields or ratio a/b, using metric (1a) in Bohr bound state $(1-1/n^2)$ theory gives

$$h\nu = hc/\lambda = [1/f(\theta_0) - 1/(n^2 f(\theta))] \sim +\{[1+(a/b)\cos\theta_0] - (1/n^2)[1+(a/b)\cos\theta\ldots]\}$$
$$\sim +(1-1/n^2) + (a/b)(\cos\theta_0 - \cos\theta/n^2) + \ldots \quad (2a)$$

for resonance to occur, with $\theta_0$ constant. Equation (2a) returns Bohr term $(1-1/n^2)$ but adds sinusoidal polarization dependent wavelength (PDW) shifts $\Delta_p$ of order a/b. H PDW shifts, exposed with (2a), cannot but be interpreted as errors $\Delta_p$ of Bohr theory. Analytically, H PDW shifts obey

$$\Delta_p \sim [(1-1/n^2) + (a/b)(\cos\theta_0 - \cos\theta/n^2) + \ldots] - (1-1/n^2) = +(a/b)(\cos\theta_0 - \cos\theta/n^2) + \ldots \quad (2b)$$
$$\Delta'_p \sim [(1-1/n^2) - (a/b)\cos\theta/n^2 + \ldots] - (1-1/n^2) = -(a/b)\cos\theta/n^2 + \ldots \quad (2c)$$

with $\Delta'_p = \Delta_p - (a/b)\cos\theta_0$, and apply for terms $(1-1/n^2)$ and energies $-1/n^2$, without affecting the sinusoidal appearance. With $1-\cos\theta = 2\sin^2(½\theta)$ and/or $1+\cos\theta = 2\cos^2(½\theta)$ for (2), non-sinusoidal effects $(a/b)/n^2$ in (2) can appear for H PDW shifts in Bohr-like H polarization theory (2a). H PDW shifts, if any, should vary as $\cos\theta/n^2$. Rewriting (2c) as $(b/a)n^2\Delta'_p \sim \cos\theta$ gives generic sinusoidal behavior. Generic polarization with phase θ is illustrated in Fig. 1a, using a cosine with asymptotes ±1. Maximum phase 2π, divided in 20 parts $2\pi/20 = \pi/10$ is centered as $-\pi, +\pi$ and scaled to $-1, +1$. The fractional angle or phase varies with $2(n-1)\pi/20$ when n goes from 1 to 21, which makes x- and y-axis commensurate. Fig. 1a for polarization needs some comments.

(i) Fig.1a exposes the generic relation between phase θ and handedness: linear polarization for $\cos\theta = 0$ (0°, parallel fields) or π (180°, antiparallel fields, with a field inversion), circular[4] polarization for poles +1 and -1, $\cos\theta = ½\pi$ (90°, say right handed, orthogonal equal fields) or $-½\pi$ (-90° or 270°, say left-handed, orthogonal equal fields) and elliptical polarization in between[5]. The phase needed to switch from left- to right-handed polarization is exactly 180°, implicit in de Broglie's relation (1h).

(ii) Sinusoidal pattern in Fig. 1a shows with laser signals in optical fibers [10-12]. When bit-compression is high (order G-THz), signals are affected by polarization. Distortions in data transmitted between Alice and Bob, due to formerly unknown polarization dependent losses [10-12], led to practical problems[6]. With Fig. 1a, observed photon (boson) polarization relates with metric definitions, entanglement, quantum computing, STR-violations, Bell-inequalities and EPR-tests using polarization of photons or neutral particles, like the kaon [10-11].

---

[4] which also requires 2orthogonal fields of equal magnitude.
[5] Complementary sine can be used for polar shifts, since polar and equatorial axes are assigned by convention.
[6] Restoring signal distortions by PMD losses or PDW shifts relies on fiber Bragg or chiral gratings [12].

(iii) In material system H, sinusoidal effects (2c) diminish with $1/n^2$. Since the smaller the size, the larger the density (compression), this situation compares with bit-density effects for photon polarization. For $1/n^2$ H theory, this gives compacted attenuated sinusoidals like in Fig. 1b-c, where $\cos(\pi n/10)/n$ and $\sin(\pi n/10)/n$ (Fig. 1b) and $\cos(\pi n/10)/n^2$ and $\sin(\pi n/10)/n^2$ (Fig. 1c) are plotted versus $1/n$. In this phenomenological H polarization model, distorted sinusoidals in Fig. 1b-c can be extrapolated to the left (expansion, low density) and even beyond $n=\infty$ or $1/n=0$, where H no longer exists (horizon problem). Extrapolation to the right (compression, high density) goes beyond ground state $n=1$, a strange consequence, discussed below. Attenuated, distorted sinusoidals in Fig. 1b-c only seem different from those in Fig. 1a but are related by Bohr packing factors $1/n$ (odd) or $1/n^2$ (even).

(iv) For resonant interactions between radiation and matter, all details in Fig. 1a for boson polarization will have to match exactly all those of fermion system H for resonance to be possible with (1e). If radiation showed left-right behavior, H must show this behavior too when its spectrum is measurable. Simplest atom H is a prototype for resonant interactions and cannot be an exception. On the contrary, the H spectrum is the simplest tool imaginable to get at polarization for the so-called electron-proton Coulomb bond, its sinusoidal effects (1a-c) and, eventually, its boson (photon) behavior.

(v) Observed line profiles (including wavelength shifts), are affected by an atom's environment. Field effects can be eliminated by extrapolation to zero field (Zeeman, Stark). Interatomic interactions of a resonating atom with identical or foreign neighbors also affect line profiles. If sinusoidal [13], these can be eliminated with extrapolation towards the difficult single atom case. Even so, their influence can be minimized using field effects for a couple of closely spaced lines, extrapolated to zero-field. This reliable procedure was used for 2 nearly degenerate lines (see the Lamb-shifts below).

(vi) With sinusoidal environmental effects accounted for, information on intra-atomic polarization is assessable. This brings in $\underline{H}$, forbidden in nature, because of charge anti-symmetry and annihilation in the Dirac-sense, although $\underline{H}$ is also electrically neutral. Theorists argue that $\underline{H}$, the right- (left-) handed version of left- (right-) handed H, is forbidden, although the two are just the natural, mutually exclusive quantum states $\pm 1$, H and $\underline{H}$, possible for a polarized H atom. The 2 atom states correspond with the 2 quantum states $\pm 1$ for bosons (radiation) or even with the generic cosine asymptotes $\pm 1$ as depicted in Fig. 1a. Instead of vetoing $\underline{H}$, one should at least have admitted a long time ago that both states H and $\underline{H}$ are theoretically possible and allowed for natural, neutral, stable composite Coulomb atom hydrogen, when polarized. If 2 quantum states $\pm 1$ for hydrogen are both stable, they must be described with an inversion along the Coulomb field axis (bipolar view), i.e. with attraction of 2 unit charges $-e_1 e_2$, say +1, inverted to $-e_2 e_1$, say $-1$. These are linear sum- and difference-states in (1a), i.e. $1 \pm m_e/m_p$ in (1c) and $1 \pm r_p/r_e$ in (1b), and result from a metric representation with polar coordinates, containing the origin. A permutation on an axis is a rotation exactly by 180° or $\pi$ for one of 2 complementary parts in H. For (2a) to give (a+b), $\cos\theta = -1$ or $\theta = \pi$ is required, while for difference (a-

b) $\cos\theta=+1$ or $\theta=0$, vector sum and difference having different orientations in space. For full resonance between H and radiation to be possible with (1e), H should be credited with some boson behavior.

(vii) If polarized H-states have always been observed from the 19th century on, these are in full resonance with radiation in the absence of (strong) external fields. If so, only the H spectrum can decide on H polarization, on boson behavior and on the fate of $\underline{H}$. Theoretically possible, plausible state $\underline{H}$ can never be forbidden, before the H spectrum is tested, without prejudice or bias, for generic H polarization effects like those in Fig. 1a.

## Testing the H spectrum for sinusoidal effects using planar Bohr $R/n^2$ fermion theory

The advantage of simple Bohr $1/n^2$ theory for fermions is that it is exclusively planar 2D (rotational freedom), without the preferential direction, needed for polarization. Bohr's circular model for the Coulomb 2 unit charge metric uses constant unit field $1/\sqrt{(\cos^2\varphi+\sin^2\varphi)}=1/\sqrt{(\cos^2\varphi+1-\cos^2\varphi)}=1/\sqrt{(1-\sin^2\varphi+\sin^2\varphi)}=1$ with constant phase 90° between sine and cosine, which gives a constant Rydberg R theory $-R/n^2$. A variable phase $\theta$ as in (1a) creates problems for circular orbits, since $1/\sqrt{[\cos^2\varphi+1-\cos^2(\varphi+\theta)]}$ is equal to 1 only for $\theta=0°$. Strange as it may seem, these difficult theoretical problems with polarization, caused by a variable not constant phase $\theta$, make Bohr $-R/n^2$ fermion theory a unique tool to quantify H behavior beyond 2D, i.e. when polarization with variable phase would show as in (2a) [2]. Theoretical level energies (in cm$^{-1}$)

$$-E_{n(2D)} = R/n^2_B \text{ cm}^{-1} \text{ or } 1/n_B = \pm\sqrt{(-E_{n(2D)}/R)} \tag{3}$$

with Bohr's $n_B$ an integer and constant R, are easily confronted with observed energies $E_{n(exp)}$, denoted in a similar way with effective or experimentally observed numbers $n_{exp}$ or

$$-E_{n(exp)} = R/n^2_{exp} \text{ cm}^{-1} \text{ or } 1/n_{exp} = \pm\sqrt{(-E_{n(exp)}/R)} \tag{4a}$$

This gives simple numerical relations

$$E_{n(2D)}/E_{n(exp)}=(n_{exp}/n_B)^2 \text{ or } \pm(n_{exp}/n_B)=\pm\sqrt{(E_{n(2D)}/E_{n(exp)})} \tag{4b}$$

Using Rydbergs $R_B$ and $R_{exp}$ in cm$^{-1}$ with integers $n_B$ in either case leads to an alternative test

$$R_{exp}=R_H(n) =(E_{n(exp)}/E_{n(2D)})R_B =E_{n(exp)}n^2_B \text{ cm}^{-1} \tag{4c}$$

with running Rydbergs $R_H(n)$ instead of constant R [2]. Differences $\Delta_n$ between $n_B$ in (3) and $n_{exp}$ in (4), not necessarily an integer if sinusoidal (2) is in operation, obey dimension-less numerical relations

$$\Delta_n=1/n_B -1/n_{exp}= (n_{exp}-n_B)/n_B n_{exp}= \pm[\sqrt{(-E_{n(2D)}/R)}-\sqrt{(-E_{n(exp)}/R)}] \tag{5a}$$

$$\Delta'_n=1/n^2_B-1/n^2_{exp}=(1/n_B-1/n_{exp})(1/n_B+1/n_{exp})=(n^2_{exp}-n^2_B)/n^2_B n^2_{exp}$$
$$=[(-E_{n(2D)}/R)-(-E_{n(exp)}/R)] \tag{5b}$$

implying that $\Delta'_n/\Delta_n\approx 2/n$. $R_{exp}$ in cm$^{-1}$ in (4c) as well as numbers $\Delta$ in (5a-b) must all be of sinusoidal type, for H polarization (2a), H boson behavior and its 2 quantum states H and $\underline{H}$ to make sense.

At the time, it was expected that results for the H spectrum using Einstein's STR-field (1d)

$$\Delta(STR)= E_{n(2D)}- \mu c^2[1/\sqrt{(1+(\alpha/n)^2)}-1] \tag{5c}$$

would be better than with constant R, the Bohr field. In reality and despite expectations, testing STR field (1d) with (5c) proves this field much less precise than naïve Bohr theory, as we show below. This summarizes the status of H bound state theories around 1920 as well as the theoretical possibilities available at the time to test Bohr theory beyond its 2D limit.

**Observed sinusoidal effects in the H line spectrum and Bohr-type polarization theory**

The H Lyman series $ns_{1/2}$ and $np_{1/2}$ are available for testing with Kelly's observed terms (errors of 0,0001 cm$^{-1}$) [14] or with Erickson's energies, according to QED calculations (errors 0,0000001 cm$^{-1}$) [15]. R, needed in (5), is 109678,7737 cm$^{-1}$ in [14] and 109678,773704 cm$^{-1}$ in [15]. Data are in Table 1. Since the consistency of H bound state QED is at stake, we use QED data [15], although Kelly data [14] give similar results. The 4$^{th}$ order fit of the 2 series [15] with $1/n$ gives respectively

$$-E_{ns}=-4,365136/n^4+5,552171/n^3+109677,586807/n^2-0,000143/n+0,000005 \text{ cm}^{-1} \quad (5d)$$
$$-E_{np}=-4,377663/n^4+5,842957/n^3+109677,585445/n^2-0,000008/n \text{ cm}^{-1} \quad (5e)$$

The fits return energies with average errors of only 7,8 kHz and 0,8 kHz respectively. Kelly terms $T_{ns}$ [14] are compatible with (5d). As an example, fitting his observed ns-terms [14] gives

$$T_{ns}=4,365740/n^4-5,553058/n^3-109677,586558/n^2+0,000133/n+109678,773744 \text{ cm}^{-1} \quad (5f)$$

which matches (5d) when comparing coefficients.

Neglecting the smaller terms in (5d) and using (4c), observed running Rydbergs for ns obey

$$R_{exp}=-E_{ns}n^2=-4,365136/n^2+5,552171/n+109677,586807 \text{ cm}^{-1} \quad (5g)$$

Tests (4c) and (5) on H symmetries or on the constancy of R lead to straightforward conclusions for the real metric needed for the so-called simple electron-proton Coulomb bond in system H.

(i) The parabolic result for running Rydbergs (4c) in Fig. 2a immediately falsifies Bohr's constant Rydberg thesis $R/n^2$ for a fermion model as it reproduces (5g). The similar result in Fig. 1 of [2][7] is repeated here for (4c) with the harmonic Rydberg (109679,35 cm$^{-1}$) included. This Rydberg is tangent to the complete series and can be considered constant in the spirit of the circular orbits of Bohr fermion 2D theory and phase 90° for sine and cosine. It touches the parabola at $n_0=\frac{1}{2}\pi$ [2] but deviates, in a sinusoidal manner, from this critical value[8]. The absence of a preferential direction in Bohr theory is visualized with horizontal lines, parallel to the x-axis in Fig. 2a, like the dashed line for n=1, the so-called H ground state at $R_1$=109678,7737 cm$^{-1}$. The classical H anchor at the extreme in Fig. 2a is obtained by removing Bohr's attenuation factor $1/n^2$ using (4c). The observed result for R is strikingly similar with that with the cosine for generic polarization in Fig. 1a, suggesting a fluctuating metric. Comparing curves, it is evident that the curves for H Rydbergs in Fig. 2a are of type (1-

---

[7] There is a typo in its Eqn. (1).
[8] This angle of 90° remains constant in planar circular models but does not show explicitly, since complementary sine and cosine are used. At first sight, retrieving this angle with a spectrum seems a logical consequence of the reality of 2 orthogonal fields and circular orbits. Yet, finding out that this angle is no longer constant in a complete series but critical instead is a true signature for H polarization [2]. This relies on deviations from 90° instead of remaining constant.

cosθ)=2sin²(½θ). Its square root implies that observed H symmetries depend on sin(½θ)√2, in agreement with polarization theory. These straightforward implications of Fig. 2a led us directly to H and justify the brief report of 2002 [2,4].

(ii) Parabolic behavior of running Rydbergs in Fig. 2a points to sinusoidal behavior, since in first order $1-\cos\theta = 1-(1-½\theta^2+...) \approx ½\theta^2$. Since this suggests boson behavior, Fig. 2a shows that a boson-like H atom appears when Coulomb attraction sets in at $1/n=0$. It reaches its maximum at $n_0=½\pi$ upon compressing the 2 H fermions (electron and proton), and then gradually diminishes. In terms of H fermion-boson symmetries, the extreme marks a transition between these 2 different H symmetries, while only H fermion symmetry can be considered as constant in the full interval, pending the choice of the asymptote ±1 to describe it. The choice of an asymptote, +1 or –1, is purely conventional but can never mean that one of them does not exist – see sine and cosine--, which also puts question marks on a veto on H. The appearance of H boson symmetry is now understandable, since it is essential for full resonance between system H and radiation (a boson structure) to be possible by virtue of (1e).

(iii) More details on H behavior beyond 2D are exposed with the curves for differences (5a)-(5b), shown in Fig. 2b,c. These are extrapolated, since it is uncertain how restricted observable domain for H, i.e. $2 \geq n \leq \infty$ or $½ \leq 1/n \geq 0$, will comply with the full domain $2\pi$ for phase θ (see further below). Compacting effects, illustrated in Fig. 1b-c, are now clearly visible with the H spectrum. Attenuation by $1/n$ or $1/n^2$ must not distract from the sinusoidal character of $\Delta_n$. In fact, the H spectrum shows that all perfectly sinusoidal curves for $1/n$ (5a) in Fig. 2b and for $1/n^2$ (5b) in Fig. 2c are consistent with parabola (5g) for $R_{exp}$, given in Fig. 2a and in Fig. 1 of [2].

Visual inspection of all Fig. 2 proves beyond any doubt that deviations from Bohr's 2D fermion model for H are sinusoidal, which confirms the existence of sinusoidal field fluctuations (2), needed for polarization to appear in natural neutral Coulomb system H. This means that the Coulomb metric is not rotationally invariant. With Fig. 1 and 2, observation [14] as well as bound state QED theory [15] both point to sinusoidal Compton-like wavelength shifts (1) in H, which led to de Broglie' equation (1f) and generic polarization angles like (1h).

To prove that these are indeed H PDW shifts, conform Fig. 1a for the polarization of light, the phase correlation of Fig. 1a must be retraced identically in the H spectrum: phase ±½π for circular and phase 0 or π for linear H polarization are both required, as argued in [2-3]. With Fig. 1-2, our earlier, rather bold conclusions on the existence of natural H-states are fully justified [2].

However, the origin of pertinent H-signatures must be retraced in bound state QED. We remind that orthogonal STR field (1d) had to be modified drastically to account for the very same H-states in Fig. 2a-b. In fact, Fig. 2d shows the result of test (5c) for STR. Highly praised STR field (1d) is worse than Bohr's: it simply proves disastrous[9]. Repulsive term in $1/n^4$ creates errors, much larger than those of simple Bohr $-R/n^2$ theory, as shown in Fig. 2d. STR errors are 12 and 27 times larger than Bohr's for respectively np and ns (not shown). To undo the damage of the STR field, large attractive corrections of order $-5,84/n^3$ cm$^{-1}$ for np and $-5,55/n^3$ cm$^{-1}$ for ns were needed (see below). Fig. 2d is illustrative for the history of theoretical physics. Perhaps theorists were misled: they were preoccupied with the great errors of the H STR-field (upper 2 closely spaced curves in Fig. 2d), which could have made them overlook the small errors of Bohr theory, wherein harmonic and quartic behavior already shows (lower 2 curves in Fig. 2d). Even with moderate spectral accuracy, the large errors of STR were clearly visible, the smaller ones of Bohr theory far less (Fig. 2d).

---

[9] This flagrant erratic behavior of STR for H (Fig. 2d) is never mentioned, it certainly was not in Einstein Year 2005 [1]

How to remove these large STR errors had a great impact on theorists but may well have distracted their attention from the real problem: H polarization and an oscillating metric[10]. It was probably not realized that Bohr theory is so reliable and powerful to disclose 3D effects beyond 2D, e.g. H polarization (Fig. 2d). This makes the history of these attractive corrections for bound states, so badly needed for the STR field (1d), quite remarkable.

**Origin of Sommerfeld's double square root equation and quartics in bound state QED**

Classical Coulomb and polarization models for composite H differ in that the first is planar (2D), due to a central field approximation, whereas the second points to out of plane effects with a bipolar view, essential to arrive at H polarization. Although Bohr's $1/n^2$ H 2D fermion theory is fairly accurate (parts in $10^7$ for terms, 300 MHz), replacing it with its also highly praised 3D Schrödinger wave mechanical version proved unsatisfactory [1], as the accuracy was no better[11]: Schrödinger returned the very same energy levels of Bohr's $1/n^2$ theory, without any correction term added.

In the early 20$^{th}$ century and before Schrödinger, physicists like Sommerfeld, Kramers, and Bohr were all occupied with the discrepancies between $-R/n^2$ theory and experiment. Theoretical corrections, inspired by Einstein's (1d), led to even greater errors than with a naïve, constant Bohr field (see Fig. 2d) and caused a great dissatisfaction with highly praised STR theory amongst notorious relativists. Even with moderate accuracy, positive as well as negative deviations from a constant Rydberg hypothesis $-R/n^2$ suggested fluctuating deviations (see Fig. 2d), e.g. blue- as well as red-shifts with respect to the level energies, predicted by Bohr's 2D model. Sommerfeld's azimuthal quantum number $\ell=n-1$ and his 2D elliptical model, useful for classifying states, did not improve the accuracy either. Unlike Schrödinger, Sommerfeld must have realized at a very early instance that only parabolic oscillatory behavior --e.g. tilted instead of planar ellipses, see (1a)-- would account for remaining discrepancies and that, at the same time, only a large attractive term $-1/n^3$ would restore the damage by STR (see Fig. 2) of great concern for relativists and theoretical physicists.

---

[10] STR is clearly violated with the H spectrum. Since H is the major source for metrology, its spectrum flaws STR immediately. Looking for violations of STR (or of Lorentz invariance) proves extremely simple and straightforward.
[11] Apparently, wave mechanical H does not improve $-R/n^2$ theory [1]. This procedure has angular dependencies *in phase* to get at resonance (and to remain soluble), instead of *out of phase* to get at polarization-states needed for full resonance.

Knowing this, Sommerfeld suggested his pupil Kratzer[12] to work on a numerical parabola [17]

$$E_p \sim (1-n_0/n)^2 \qquad (6)$$

where $n_0$ is a critical internal quantum number for H, missing in Bohr-Schrödinger theories. Although at the time, Kratzer's potential (6) was considered typical for oscillations between atoms in molecules[13] (see Appendix), Sommerfeld knew it could improve the precision of a new theory, in which the Bohr-Schrödinger $1/n^2$ term had to remain. His own quantum number $\ell=n-1$ and

$$\alpha = e^2/\hbar c = 1/137,035999\ldots \qquad (7)$$

his numerical field scale factor (the fine structure constant, first referred to in 1915), should be conserved too. The Sommerfeld-Kratzer connection is now easily understood. Sommerfeld's number $\ell=n-1$ is connected with Kratzer's potential (6): with $n_0=\ell$, (6) is degenerate with Bohr's $1/n^2=(1-\ell/n)^2=[1-(n-1)/n)]^2=(1-1+1/n)^2=1/n^2$. But for $\ell=0$ (circular orbits), the effect of (6) with integer n would vanish identically. Even with constant $n_0$, (6) gives results equivalent to Bohr's, for any intermediary asymptote (Rydberg), virtual or not, we would choose. Putting $R/n^2=A(1-n_0/n)^2$, always returns an exact numerical relation between a scaled asymptote $\sqrt{(R/A)}=n-n_0$, without loss of precision. The procedure (not shown) is easily verified with a plot of terms or levels versus $n_0/n$ or versus $(n_0/n-1)$, with $n_0$ integer or not. The linear $1/n$ procedure is exemplified with (5a) and Fig. 2b. With this evidence in mind, Sommerfeld also knew too well that the Einstein-STR expansion for H on the basis of (1d), used for (5c)

$$-E_n/\mu c^2 = [1/f_{STR}(n)-1] = 1/\sqrt{(1+\alpha^2/n^2)}-1 = 1-\tfrac{1}{2}\alpha^2/n^2+\tfrac{3}{8}\alpha^4/n^4-\ldots-1 = (-\tfrac{1}{2}\alpha^2/n^2)(1-\tfrac{3}{4}\alpha^2/n^2-\ldots) \qquad (8)$$

can never give parabolic behavior of type (6). Term $+\tfrac{3}{8}\alpha^4/n^4$ in (8) may be small indeed and of the required magnitude, it remains exclusively repulsive. For H, it creates rather than solves problems for Bohr $1/n^2$ theory, as shown in Fig. 2d. Sommerfeld associated the more visible Kratzer parabola (6) with higher order attractive terms in an STR expansion, to arrive at terms of mixed type

$$E_p \sim (\alpha^4/n^2)(1-n_0/n)^2 \qquad (9)$$

intimately connected also with his quantum number $\ell$ as shown above.

STR (8) had to be modified drastically, while Bohr, Kratzer (6) and Einstein STR (8) theories as well as his quantum number $\ell$ had to remain. To get a STR-type field in line with observation was a matter of mathematical skills, like Sommerfeld's.

These led him to a remarkable, ingenious double square root[14] solution with reduced mass[15] $\mu$, i.e.

---

[12] Rigden [16] does not mention that Kratzer was Sommerfeld's pupil. Kratzer's potential $V(r)=-e^2/r +b/r^2$ and $V(r_0)=-\tfrac{1}{2}e^2/r_0$ figures at length in Sommerfeld's famous monograph [18]. It is probably the most underestimated potential in physics and chemistry [19](see also Appendix). A molecular Kratzer potential is universal. It is superior to Morse's and accounts smoothly for lower order spectroscopic constants of 300 diatomics $X_2$ [19-20]. Kratzer's potential (6) refers to the 19th century ionic Coulomb $X^+X^-$, not to the covalent asymptote XX. This seems to favor old-fashioned classical Coulomb ionic bonding $-e^2/r$ but, in reality, it gives away atom-antiatom or X<u>X</u> bonding [1,20], although a ban on <u>H</u> implies a ban on H<u>H</u> and on H<u>H</u>-oscillations [21]. The conventional argument, often used, against X<u>X</u> bonding is that it contradicts mainstream physics and chemistry!
[13] Sommerfeld obviously made oscillations (bosons) interfere with rotations (fermions): this makes him the pioneer of supersymmetry (SUSY, MSSM), although, at the time, the fermion-boson symmetry concept was not used.
[14] More than a century ago, Ramanujan, once at Trinity, found a connection between double square roots and quartics [22].

$$E_{n,j} = \mu c^2[f(n,j)-1] = \mu c^2 f(n,j) - \mu c^2 \tag{10}$$

$$f(n,j) = \{1 + \alpha^2/[\sqrt{((j+\tfrac{1}{2})^2 - \alpha^2)} + n - j - \tfrac{1}{2}]^2\}^{-\tfrac{1}{2}} \tag{11}$$

This new STR-like field factor $f(n,j)$ in (11), today textbook material and once considered as the latest real cornerstone of theoretical physics, is similar to $f_{STR}(n)$ in (1d) or (8) but different. Its validity, if confirmed by the H spectrum, would immediately flaw original Einstein STR equation (1d) or (8), as evident with Fig. 2d. Also, $j$ is the total angular quantum number, with values $j = +\tfrac{1}{2}$ for $\ell = 0$ and $j = \ell \pm \tfrac{1}{2}$ for $\ell \neq 0$, where $\ell = n-1$ is Sommerfeld's number. It is retrieved exactly in (10)-(11) for atom states obeying $j + \tfrac{1}{2} = 1$, for which $n - j - \tfrac{1}{2} = n - (j + \tfrac{1}{2}) = n - 1 = \ell$. Looking at Fig. 2d, these derivations place the famous double square root equation (10)-(11) of modern bound state QED in a different context: it was simply needed by Sommerfeld to fit his azimuthal quantum number and oscillatory (boson) behavior (6) into a very badly performing relativistic Bohr-Einstein (fermion) rotator (8) because of sinusoidal errors remaining for H with Bohr's planar $1/n^2$ theory[16].

In fact, the difference between (10) and (8) may be subtle, the connection with $\ell$ and Kratzer's (6) is obvious. For $j + \tfrac{1}{2} = 1$ with $\ell = 0$ and $\ell = 1$, terms between square brackets in (11) simplify as

$$[\sqrt{((j+\tfrac{1}{2})^2-\alpha^2)}+n-j-\tfrac{1}{2}]^2 = [\sqrt{(1-\alpha^2)}+(n-1)]^2 = [\sqrt{(1-\alpha^2)}+\ell]^2$$

$$= (n-\tfrac{1}{2}\alpha^2)^2 \approx n^2 - n\alpha^2 = n^2(1-\alpha^2/n) \tag{12}$$

$$f(n,j+\tfrac{1}{2}) = \{1 + (\alpha^2/n^2)/(1-\alpha^2/n)\}^{-\tfrac{1}{2}} = \{1 + (\alpha^2/n^2)(1+\alpha^2/n)\}^{-\tfrac{1}{2}} \tag{13}$$

Expanding these like in (8) generates a parabola for H of the required Kratzer form (6), since

$$E_{n,j+\tfrac{1}{2}}/\mu c^2 = [f(n,j)-1] = \{1 + (\alpha^2/n^2)(1+\alpha^2/n)\}^{-\tfrac{1}{2}} - 1$$

$$= -\tfrac{1}{2}(\alpha^2/n^2)(1+\alpha^2/n) + \tfrac{3}{8}(\alpha^4/n^4)(1+\alpha^2/n)^2 - \ldots \approx -\tfrac{1}{2}\alpha^2/n^2 - \tfrac{1}{2}\alpha^4/n^3 + \tfrac{3}{8}\alpha^4/n^4 - \ldots$$

$$= (-\tfrac{1}{2}\alpha^2/n^2)[1+\alpha^2(1/n-\tfrac{3}{4}/n^2)-\ldots] \tag{14}$$

In first order, the final parabolic compact energy level equation for H becomes

$$E_{n,j+\tfrac{1}{2}} = (-\tfrac{1}{2}\mu\alpha^2 c^2/n^2)[1+\alpha^2(1/n-\tfrac{3}{4}/n^2)-\ldots]$$

$$= (-R/n^2)[1+\alpha^2(1/n-\tfrac{3}{4}/n^2)-\ldots] \tag{15}$$

much more accurate than Bohr's leading 2D $-R/n^2$ term, still preserved. It is too easily forgotten that bound state formula (15) is due to Sommerfeld, not to Dirac, and that it remains at the basis of modern bound state QED [24].

With (15) and numerical dimension-less ratio $-E_{n,\tfrac{1}{2}}/R$, errors $\delta_n$ for Bohr theory like $\Delta_n$ in (5) are easily quantified by a clearly visible quartic (in cm$^{-1}$)

$$.\quad \delta_n = -E_{n,\tfrac{1}{2}} - \tfrac{1}{2}\mu\alpha^2 c^2/n^2 = \alpha^2 R(1/n - \tfrac{3}{4}/n^2)/n^2 \text{ cm}^{-1} \tag{16}$$

for H $np_{\tfrac{1}{2}}$ states. This justifies equations (4c) and (5a-b), the curves in Fig. 2 and in [1,3] but also gives the attractive term $-\alpha^2 R/n^3 = -5,84/n^3$ cm$^{-1}$ needed to remove the STR errors in Fig. 2d (see above).

---

[15] It is strange that (10) uses $\mu$, although at $n = \infty$, an electron with mass $m_e$ instead of $\mu$ is set free. There is no reason to use reduced mass in STR based (10), as remarked by Cagnac et al. with some irony [23]. They write that the only justification (sic)[used by Cagnac] to use reduced mass $\mu$ in (10) is that it makes this equation consistent with experiment [23] (see below).

[16] Einstein's original non-sinusoidal orthogonal STR (1d) can be modified for bound H states with an amended form like $f'(\alpha/n) = [1 + (\alpha/n)^2/\sqrt{(1-\alpha^2/n)^2}]^{-\tfrac{1}{2}} = [1 + \alpha^2(1+\alpha^2/n+\ldots)/n^2]^{-\tfrac{1}{2}}$, the parabolic Sommerfeld variant hidden in (14)-(15). This suggests that the harmonic H field would derive from a series expansion of a series.

Sommerfeld-QED quartic (16) transforms in numerical variants $\delta_n/R\alpha^2$, $n\delta_n/R\alpha^2$ and $n^2\delta_n/R\alpha^2$, shown in Fig. 3. The Mexican hat shape shows only with extrapolation as in Fig. 2. Using (16), cubic $n\delta_n/R\alpha^2$ is harmonic between a parabola with $n^0$ and a quartic with $n^2$, since $n\delta_n/R\alpha^2 = \sqrt{[(\delta_n/R\alpha^2)(n^2\delta_n/R\alpha^2)]}$ and explains why all curves converge to the same critical points (Fig. 3). With Sommerfeld's, unjustly called Dirac's equations (14)-(15), running Rydbergs (4c) are parabolic

$$R(n) \approx R[1+\alpha^2(1/n - \tfrac{3}{4}/n^2)] \tag{17}$$

as argued in [2]. With R=109677,5854 cm$^{-1}$ as in (5e) and (7) for $np_{1/2}$, Rydbergs and energies become

$$R(n) \approx 109677{,}5854 + 5{,}8405/n - 4{,}3805/n^2 \text{ cm}^{-1} \tag{18a}$$

$$-E_{n,1} = -4{,}3805/n^4 + 5{,}8405/n^3 + 109677{,}5854/n^2 \text{ cm}^{-1} \tag{18b}$$

in perfect agreement with fit (5e). Before 1947, this impressive precision for H got Sommerfeld's, not Dirac's, bound state theory (15) even the status of an absolute theory. This had major consequences for metrology, based on the Rydberg $m_e\alpha^2c^2$, still in vigor today [23]. It is evident that Sommerfeld[17] is responsible for this status, since he succeeded in safeguarding Einstein's highly praised STR formalism for bound states. Unfortunately, the obvious link with H boson behavior and its possible impact for H polarization was never made. The irony is that Sommerfeld almost unwillingly proved that simple oscillating H fields (1a-c), say a cosine law for the Coulomb metric, are far better than Einstein's original orthogonal STR field, which, by definition, can never cope with polarization (see above).

**Hidden Kratzer oscillatory potential and harmonic Rydbergs in fermion-boson system H**

With this pragmatic origin of the famous double square root equation (10), Kratzer's potential (6), showing in (10)-(11), retains its classical (boson) implications even when superimposed on fermionic H (see Appendix). Apart from factor $\alpha^2$ and a shift, Sommerfeld-Kratzer potentials (6) and (15) give

$$-(1-n_0/n)^2 = -1 + 2n_0/n - n_0^2/n^2 \sim +1/n - \tfrac{3}{4}n^2 \tag{19}$$

The numerical Kratzer parabola (6) hidden in H $np_{1/2}$ refers to a critical n-value $n_0=3/2$, half integer and constant, instead of running $\ell=n-1$. The Kratzer potential needed for $np_{1/2}$-states becomes

$$E_p = -(1/3)(1-n_0/n)^2 = -(1/\sqrt{3} - \tfrac{1}{2}\sqrt{3}/n)^2 = -1/3 + 1/n - \tfrac{3}{4}n^2 \tag{20}$$

which, in turn, refers to a different asymptote $E'_p$, shifted numerically by

$$E'_p = -(1/3)(1-n_0/n)^2 + 1/3 = -1/3(1-1) + 1/n - \tfrac{3}{4}n^2 = +1/n - \tfrac{3}{4}n^2 \tag{21a}$$

This asymptote shift can be dealt with using classical physics and remains mathematically exact using a virtual particle-antiparticle pair asymptote difference (1-1)/3 in (**21a**). This freedom of asymptote for H is connected with the form of Kratzer's oscillatory potential (6), i.e. with or without constant asymptote shift, governed by $n_0$ in (6).

---

[17] It is too easily forgotten also that Sommerfeld's work was much admired. For 1901-1950, this got him the highest number of nominations (81) for a Nobel prize but never received it (Bohr got 20 nominations) [25]. A. Kratzer and E. Fues but also P. Debye, W. Pauli, W. Heisenberg and H. Bethe belonged to Sommerfeld's *school* (http://www.lrz-muenchen.de/~sommerfeld).

An important aspect of asymptote shifts, never mentioned in bound state QED and by NIST [2], is the appearance of harmonic Rydbergs $R_{harm}$. For H $ns_{1/2}$, $R_{harm}$=109679,3522 cm$^{-1}$, while for n=1, the ground state, $R_1$=109678,7737 cm$^{-1}$ [2] (see also below). For $np_{1/2}$ with (21), shift $\alpha^2 R/3$ =1,94 cm$^{-1}$ gives R'$_{harm}$109677,58+1,94=109679,52 cm$^{-1}$, whereas for its ground state at n=1, shift $\alpha^2 R/4$ gives 109679,04 cm$^{-1}$. The R-parabola in Fig. 2a learns that underlying linear H field of type a-b/n or sin(½θ)√2 (see around Fig. 2), never discussed in bound state QED, will have to be understood from first principles in ab initio H polarization theory [26]. For np, R'$_{harm}$ is shifted by $\alpha^2/3$, which is important to understand observed distorted quartics, extracted from the lines. Fig. 4 shows that the invisible theoretical harmonic quartic $\delta_n$, scaled with $\alpha^2$, or

$$\delta_n/(\alpha^2)=(1/3)(1-1,5/n)^2/n^2 \tag{21b}$$

is symmetrical and critical at n=3, whereas the clearly visible observed anharmonic quartic

$$(1/3)[(1-1,5/n)^2-1]/n^2 \quad \text{(giving observed } +1/n^3-\tfrac{3}{4}/n^4 \text{ after inversion)} \tag{21c}$$

is not only asymmetrical but also its shape is markedly different. This also shows how linear asymptote shifts can affect the shape of the quartics, hidden in the H line spectrum, as argued above. Exposing harmony with parabolas relies on scaling effects on coefficients and the value of asymptotes. Forms $(a-b/n)^2=a^2[1-(b/a)/n]^2=b^2[(a/b)-1/n]^2=2ab[\sqrt{(½a/b)}-\sqrt{(½b/a)}/n]^2$ are all equivalent. Rewriting the latter with ratio r=(b/a) gives harmonic relation $2(ar)^2(½/r^2-1/n+½r^2/n^2)$ as it is observed (21c).

However, exposing details on H symmetries also suffers from the natural limits imposed on observation. Due to (1e), observable lines are limited to the quantum domain between n=2 and n=∞ (or between 1/n=0,5 and 1/n=0) as indicated with the 2 vertical lines in Fig. 4.

With this natural limitation of (1e), it is apparent that, to expose all the details of harmonic H-behavior (boson symmetry), extrapolation[18] beyond this observable region is essential [3], as argued also above. These details all derive from Sommerfeld's decision to introduce Kratzer's oscillator potential (6) in a bound state H theory. This important Sommerfeld-Kratzer connection on classical rotator-oscillator (now fermion-boson) physics is never mentioned in the history of H bound state QED. On the contrary, QED, the QFT for the electromagnetic field, is connected almost exclusively with Dirac[19]. Sommerfeld's double square root equation (10) cannot but refer to Mexican hat curves for bound Coulomb H states, whereas the link between quartics and chirality was known already with 19$^{th}$ century chemistry [27-28].

Why all this was nevertheless persistently overlooked in QED, is difficult to understand: it was well known that radiation cannot but act like a boson system, which exhibits chiral behavior. This makes oscillator contributions (6) for resonant polarized H-states quite plausible, the more since also polarization angles or phases were already available with de Broglie's (1f).

---

[18] Bohr theory is based on extrapolation too to get at ground state n=1 in a wavenumber or 1/λ view. This avoids the infinity with (1e) for state n=1 in a wavelength or λ view (Angstrom), the spectral unit in vigor in the 19$^{th}$ century.
[19] In the latest review on H bound state QED [24], Dirac is mentioned 116 times, Sommerfeld and Kratzer not once.

At this stage, it is not yet evident to correlate quartic (16) with chiral behavior of polarized H, since quartics apply for most transitions in 2 level systems like order-disorder transitions (see below).
A direct link between H spectrum and H polarization like in Fig. 1a, can only be provided if and only if critical de Broglie polarization angles 90° and 180° are retrieved exactly from the H line spectrum.

**H polarization and H chirality: PDW shifts and spectral signatures for H̲**

Quantitative signatures for H polarization, and hence for natural H̲, only appear with the Lyman $ns_{1/2}$ series. With $j+½=1$, these states also comply with Sommerfeld's secondary quantum condition $\ell=n-1$. If (15) were the result of an absolute theory, as commonly believed at the time, $ns_{1/2}$ should be degenerate with $np_{1/2}$ and both should obey (18).
This theoretically predicted degeneracy was, however, flawed for interval $2S_{1/2}$-$2P_{1/2}$ by Lamb and Retherford [29]: the terms differ by more than 1000 MHz or 0,035 cm$^{-1}$, in line with what many other physicists already suspected earlier [29]. Immediately after its publication in 1947, the shift caused a great turmoil in theoretical physics, since the almost sacred, so-called absolute Sommerfeld-Dirac formula (10) and (15) proved wrong. The shift initiated the search for a new physics and later, following Bethe's first explanation for the shift [30], led to modern quantum field theories and eventually the Standard Model as we know it today. How important the Lamb shift may have been for theoretical physics [16,24], it is even more important for the fate of natural H̲, although this was never mentioned, until 2002 [2]. Its importance becomes apparent with $ns_{1/2}$ fit (5d)

$$-E_{ns_{1/2}}=-4,3651/n^4+5,5522/n^3+109677,5868/n^2 \text{ cm}^{-1} \quad (22)$$

with slightly different higher order coefficients than in (18) [2]. These small differences correspond with the Lamb-shifts between the 2 series.
Instead of critical $n_0=3/2$ for $np_{1/2}$, the derivative of (22) leads to a critical $n_0$ for its Kratzer potential (6) and for its running Rydbergs $n^2E_{ns_{1/2}}$ at [2]

$$n_0=1,572 \approx ½\pi \quad (23)$$

Reminding polarization Fig. 1a, phase (23), typical for orthogonal models with complementary sine and cosine of the same angle like Bohr's, is nevertheless also the critical angle for circular H polarization, pointing towards left- and right-handed configurations H and H̲.
The Sommerfeld-Kratzer super-symmetric quartic correction for H Lyman $ns_{1/2}$ gives for shifts (2c)

$$\text{HPDW shifts} \sim (1/\sqrt{\pi}-½\sqrt{\pi}/n)^2/n^2 = (1/\pi)(1-½\pi/n)^2/n^2 \quad (24)$$

instead of quartic (16) and parabola in (20) for $np_{1/2}$. Classically, and apart from trivial $n_0=\infty$, this quartic for H PDW shifts is not only critical for $n_0=½\pi$ in (23) but also for $n'_0=\pi$, instead of 3/2 and 3 for $np_{1/2}$. This shift in critical behavior for $ns_{1/2}$ provides with the 2 ultimate spectral signatures for H polarization and for natural H̲. With quartic (24), not only ½π in (23) for circular H polarization but also critical angle of π or 180° for linear H polarization shows up unambiguously as

$$n'_0=\pi \quad (25)$$

as with de Broglie's theoretical polarization angles (1f).

This finally proves that observed H wavelength shifts, i.e. errors with respect to Bohr 2D fermion $1/n^2$ theory, are polarization dependent wavelength (PDW) shifts in important H Lyman's $ns_{1/2}$-series. Angle (25) for a simple Coulomb H bond provides with an absolute, generic signature for an internal permutation (by rotation), i.e. an inversion on the axis from →, say +1, to ←, say −1. Without (strong) external fields or with data extrapolated to zero-field (Lamb-Retherford) and with oscillations projected on a Coulomb field axis, inversion (25), implicit with (1a), i.e. H acting as a half wavelength plate in de Broglie theory (1f), must take place between the 2 charges, conventionally assigned to electron and proton, which formally become anti-electron (positron) and antiproton. If state +1 stands for a natural, stable Coulomb H-state with attraction $-e_1e_2/r$, inverse −1 must stand for a natural stable Coulomb $\underline{H}$-state with attraction $-e_2e_1/\underline{r}$.

Nevertheless, natural $\underline{H}$ is still forbidden by the physics establishment, while bound state QED uses the very same Sommerfeld quartic generating field equation (10). Failing to see critical H phase[20] ½π and π, hidden in de Broglie's equation, at an earlier instance had devastating consequences for physics [1]. Apart from an unjust ban on natural $\underline{H}$, due to Dirac theory, many other important issues are affected: CPT, WEP, Lorentz invariance, existence of antimatter, cosmology, dark matter, quintessence, hole-theory, matter-antimatter asymmetry, Big Bang, H$\underline{H}$ bonding, H$\underline{H}$ oscillations, geometric phase, phase transitions, entanglement…. [1,21] (see further below).

**Results**

*(i) Precision*

To test the reliability of and the precision behind the π-dependence for the $ns_{1/2}$ series, compared with that on 3 for the $np_{1/2}$-series, we verify that, with (20) and (24), parabolic Rydberg ΔR corrections are

$$\Delta R(np) = \alpha^2 R(1/\sqrt{3}-\tfrac{1}{2}\sqrt{3}/n)^2 = +1,946827-5,840480/n+4,380360/n^2 \text{ cm}^{-1} \quad (26a)$$

$$\Delta R(ns) = \alpha^2 R(3/\pi)(1/\sqrt{\pi}-\tfrac{1}{2}\sqrt{\pi}/n)^2 = +1,775293-5,577248/n+4,380360/n^2 \text{ cm}^{-1} \quad (26b)$$

with $\alpha^2 R=109677,58545/137,035999^2=5,840480$ cm$^{-1}$. In as far as a number 3-based n-theory (26a) is of first principles character, a rescaled version with π instead of 3 is a derived first principle's theory also, as evident with (26b). The theoretical Rydbergs are plotted versus the observed ones ***in*** Fig. 5. Linear fits give

$$R(np)=109679,533968 -1,001024\Delta R(np) \approx 109679,533968 -\Delta R(np) \quad (27a)$$

$$R(ns)=109679,353174 -0,995566\Delta R(ns) \approx 109679,353174 -\Delta R(ns) \quad (27b)$$

The 2 harmonic Rydbergs differ by 109679,53-109679,35=0,18. With (26ab), level energies become

$$-E(np)=(109679,533968-\Delta R(np))/n^2=109677,5854/n^2+5,840480/n^3-4,380360/n^4 \text{ cm}^{-1} \quad (28a)$$

$$-E(ns)=(109679,353174-\Delta R(ns))/n^2=109677,5854/n^2+5,577248/n^3-4,380360/n^4 \text{ cm}^{-1} \quad (28a)$$

---

[20] Herbst [31] uses critical ½π as a Friedrichs extension for a one-particle Hamiltonian in Coulomb potential $V(r)=-n_0/n$, with lower bounds $E \geq mc^2\sqrt{[1-(\tfrac{1}{2}\pi n_0)^2]}$ in the range $0 \leq n_0 < 2/\pi$ (see [32] for a discussion).

which give an absolute error of 2,04 and 0,64 MHz for ns and np respectively. Using the fits in (27), the errors reduce to 0,24 and 0,18 MHz (some 0,000007 cm$^{-1}$) as shown in Table 2. Part of the remaining small errors is due to adaptations in constants and conversion factors since [15].

To the best of my knowledge, this is by far the simplest accurate *one-parameter* theory ever to account for observed Lamb-shifts from first principles. The explanation for the Lamb shift in standard QED is one of the most complex exercises in theoretical physics. This complexity shows when it comes to calculate H line intensities, intimately connected with H polarization. Bound sate QED is not a closed but an open theory, which adds to the ambiguity surrounding its labyrinth of terms created with the expansions for (10).

The rationale behind our simple solution for Lamb-shifts is H polarization, persistently but unjustly overlooked in QED/QFT [1].

*(ii) The 21 cm H line and natural H*

The two harmonic quartics for the Lyman $ns_{1/2}$- and $np_{1/2}$-series are shown in Fig. 6 but an important observed internal anchor for H is added, the 21 cm line. The observed hyperfine splitting of ground state $1S_{1/2}$ 1420,4058 MHz (21,1061133 cm) is 0,04737964 cm$^{-1}$, as in Fig. 6 (horizontal line). This usually faint line, important for cosmology and for mapping the Universe, is easily measured, since H is so dominant in the Universe (CBR). It is easily computed in QED with electron and proton magnetic moments. However, QED cannot place this line, important for H-related polarization effects in the Universe, within the context of a polarized H atom, where it really belongs, since QED is internally inconsistent exactly where it matters: H polarization.

Barrier-heights in its Mexican hat curves reflect all of the basic symmetries in H. Fig. 6 shows that the famous 21 cm line is exactly in between barriers in $ns_{1/2}$- and $np_{1/2}$-quartics, with respective heights 0,044649 at n=π and 0,054273 cm$^{-1}$ at n=3. This novel but unexpected result again calls for a generic H polarization or CSB theory, with due respect for this 21 cm line [26].

*(iii) Role of π in quantum rules and de Broglie quantization*

Kratzer's potential allows rotations and/or oscillations (see Appendix), while Bohr's theory only gives rotations. Their difference must show in quantum rules. In Bohr's quantum hypothesis

$$(mvr)_B = n\hbar \qquad (29a)$$

Planck's quantum of action h is scaled with the circumference of a unit circle 2π, giving ℏ. Since quantum numbers are projections on the axis, Bohr's hypothesis hides the possibility that, for linear H, oscillations along an axis obey a rescaled quantum hypothesis of type

$$(mvr)_{SK} = (k/2\pi)h \qquad (29b)$$

which could be called the Sommerfeld-Kratzer quantum hypothesis, if it were not already available with de Broglie's quantum recipe for fields (1e). Obviously, k is related to Bohr's n by scaling only or

$$k = n/2\pi \qquad (29c)$$

in line with de Broglie's (1e) and appears with Rydbergs $R_{SK}$ or H-state energies scaled accordingly. Results (29a) and (29b) warn against improper use of ℏ in so-called absolute theories with ℏ=c=1. For a de Broglie state with n=π, i.e. for the classical anchor of H in the ns-series, k=½ is a constant half integer quantum number, not only typical for H harmony but also for H, acting as a half wavelength plate. In second order, this value is also given away by half integer spin for fermions and is exposed as such by atom H, when its spectrum is measured in (strong) external fields.

If H harmony were expressed with n=π or k=½, harmonic scale factor $1/\pi^2$ affects conventional H views by a factor of some 1/10. Conventional recoil being 60 cm$^{-1}$ (109677,58/1836 cm$^{-1}$), recoil in a de Broglie harmonic state is only 6 cm$^{-1}$, close to $\alpha^2 R$=5,84 cm$^{-1}$ in (26a). We return to all this in [26].

*(iv) Classical Van der Waals-Maxwell-type phase transition in atom H*

Inversion within system H at critical n=π implies that a phase transition occurs from state +1 or phase θ=0 to state –1 or phase θ=π, the rationale behind the H quartics or Mexican hat curves above.

If not bound to chirality, double well curves give away an internal phase transition in a 2 level quantum system (order-disorder, state of aggregation…). These H potential energy curves (PECs) are obtained with energy differences, i.e. H PDW shifts, plotted versus 1/n (see Fig. 1 in [3]). Taking energy differences from the Bohr ground state n=1 gives the slightly distorted Mexican hat curve of Fig. 7a, instead of the harmonic quartic for H PDW shifts with the harmonic Rydberg [3]. To illustrate the effect of asymptote of R-shifts, curves for intermediate and more extreme R>$R_{harm}$ are also shown. Obviously, the well at the left, near to the origin at 1/n=0, represents the conventional Bohr H-atom e$^-$,p$^+$ with $r_H=r_e+r_p > r_e$ if field (1b) is taken as reference: it is the matter well. The second well at the right, appearing for smaller n or larger 1/n stands for the charge-inverted H-system e$^+$,p$^-$ with $r_H=r_e-r_p$ < $r_e$: it is the antimatter or H-well.

However, when these same small energy differences, i.e. H PDW shifts, are plotted versus n as in Fig. 7b, the resulting curves not only loose their typical harmonic quartic shape. Quite surprisingly, typical Vander Waals-Maxwell patterns show up for all curves in the quantum interval n=1 to n=∞. The upper continuous curve for ideal behavior, i.e. the ideal gas law of the 19$^{th}$ century, refers to the ideal Coulomb law in Bohr's version –R/n$^2$ for the quantum world. The relevance of Fig. 7b is improved with the 21 cm H line included as reference. The inverted Van der Waals n-view in Fig. 7b calls for a classical explanation of the H system with density fluctuations upon compressing a neutral 2 unit-charge system by decreasing their separation [5, 33], accompanied by an inversion from e$^-$,p$^+$ to e$^+$,p$^-$. Mass or density fluctuations along the radial Coulomb field axis are usually not considered for H but readily appear with reduced mass, instead of total mass (see Appendix). Moreover, the procedure applied to go from the H quartic in Fig. 7a to the Van der Waals-type curve in Fig. 7b is readily inverted. To get at a classical Van der Waals curve in a P,V diagram, pressure data P are plotted versus volume V, as in Fig. 7b. The classical Maxwell, double well or Mexican hat curve immediately shows when plotting the same pressure data versus 1/V instead, as in Fig. 7a [33].

Since critical points in the H spectrum in whatever analytical relation refer to relative contributions of fermion and boson symmetries in H, the puzzling result in Fig. 7b must relate to atomic BECs (Bose-Einstein condensates) [33]. With this phenomenological analysis, a striking one-by-one correlation appears between (i) macroscopic 19th century Van der Waals-Maxwell behavior of neutral systems with 2 phases (gas, liquid or water, steam for $H_2O$) and their classical phase transition and (ii) the microscopic phase transition between to different phases H and $\underline{H}$ of the same two level quantum system, natural and neutral atomic species hydrogen [33].

Classically, Fig. 7b means that at long range (low density, positive pressure domain) for n>π (harmonic Rydberg) or for n>5,5 (Bohr Rydberg), the hydrogen system H behaves like a conventional electron-proton system like Bohr's. At short range (high density, negative pressure domain) for n<π (harmonic Rydberg) or for n<5,5 (Bohr Rydberg), it behaves like a positron-antiproton system, i.e. an antihydrogen system $\underline{H}$. In cosmological models, negative pressure domains are associated with dark matter and a non-Coulomb equation of state (of Van der Waals type) [35]. Quite surprisingly, the H spectrum brings in quintessence and the cosmological constant for the Universe. Then, Fig. 7b may be illustrative also to understand dark matter through neutral antimatter, a surprising, unexpected link but beyond the scope of the present analysis of internal H-symmetries.

**Discussion**

$\underline{H}$-signatures (23) and (25), theoretically allowed by de Broglie's standing wave equation, were already overlooked in the earliest days of quantum field theory and especially in the aftermath of the Lamb-shift. If the H line spectrum were interpreted along these lines, a theoretical ban on natural $\underline{H}$ and on H$\underline{H}$ would never have appeared. In QFT, handedness or helicity is connected with particle spin ±½ in a dynamic approach. Yet, with quantum condition j+½=1 for equation (10), dynamic effects of half integer spin vanish, which means that parabolic, sinusoidal variations (23) for the $ns_{½}$ series of natural, neutral and stable species H can only be accounted for with a generic H polarization or CSB theory, classically and quantitatively in line with cosine law (1) for 2 internal harmonic H fields. This model underlies our Bohr-like H polarization theory (2a) in beta-version [26]. To the best of my knowledge, no ab initio H polarization theory exists today (see Introduction). In a first principles theory, field ratio a/b and fractional polarization angle θ in (1a-c) must be identified analytically [26].

The use of more or less constrained Bohr and Kratzer potentials reduces to physical differences between mathematically equivalent descriptions of circles (see Appendix). In Bohr's standard central field approximation, a circle is described with a freely rotating radius r (0,+1) and a phase of 90° between 2 orthogonal fields (radial and angular fields or static and dynamic fields). In the mathematically equivalent bipolar view, the circle is described with a diameter 2r and two poles (+1,-1) with a phase of 180° between 2 parallel or antiparallel fields (linear 2 field case, electron- and proton Coulomb fields), as suggested with de Broglie's equation (1g). In [26], we analyze 2 internal sub-field models: (i) cosine law (1a), which implies sinusoidal fluctuations between field sum and field difference, both having different directions in space and (ii) its linear bipolar variant $a\cos\alpha+b\cos\beta$, for which $a\sin\alpha=b\sin\beta$ and $(\alpha+\beta)=\pi-\theta$, which implies fluctuations of the origin instead (vacuum fluctuations). These equations for the metric appear for all harmonic periodic motions, governed by $\cos\alpha+\sin\beta$. Equivalent numerical field equations (1a-c) for H, add to the confusion about the improper use of reduced mass in bound state QED (10), as remarked above, in the Appendix and referred to in [34]. In fact, for harmonic reduced mass to appear in bound state theories on a gravitational basis, total mass $m_H$ looses its scalar behavior (see Appendix). The rationale is that recoil transforms exactly from multiplicative to additive in

$$1/(1+m_e/m_p) \equiv 1 - m_e/m_H \qquad (30a)$$

which is important for field sum and difference in (1c) [26]. The derived symmetry behind (30a) for reduced mass leads to

$$(1+m_e/m_H)/(1 - m_e/m_H) = 1,0011\ldots \qquad (30b)$$

numerically close the anomalous electron mass [2]. In this form, H line splitting in function of

$$1 \pm m_e/m_H \qquad (31)$$

would make total H mass $m_H$ indirectly responsible for the breaking of left-right symmetry for boson system H and, by extension, for a 2 unit charge Coulomb bond between 2 fermions. Not surprisingly, this is consistent with observation as argued before on a phenomenological basis [34]. An objection to (31) could be that recoil $(m_e/m_p)R$ gives about 60 cm$^{-1}$, too large to account for the H observed oscillations in QED of order $\alpha^2 R \approx 6$ cm$^{-1}$ or 10 times less. This conventional argument on recoil is, however, deceptive for harmonic H states, as shown in the foregoing paragraph.

An argument in favor of recoil[21] as a symmetry breaker is that its oscillations with frequency $\omega$, instead of rotations with angular frequency $\omega/2\pi$, compares well with a rescaled fine structure constant, although this effect is not manifest in de Broglie's original standing wave equation (1e). Confronting Coulomb and radiative fields $e^2/r$ and $h\nu=hc/\lambda=2\pi\hbar c/\lambda=2\pi(e^2/\alpha)/\lambda$ gives

$$\lambda/r=2\pi(e^2/\alpha)/e^2=2\pi/\alpha= 6{,}28\cdot 137= 861 \qquad (32)$$

the hidden scale factor behind de Broglie's variant (1f). We notice that twice recoil is $(1836/2)=918$ gives a difference of about 57 with (32) but also that $\alpha/2\pi=r/\lambda$ in (32) appears as the leading term in the Schwinger expansion for the so-called anomalous electron mass [36]. Corrections for anomalous electron mass are also used in bound state QED [24], which strengthens the confusion about the real role, played by recoil. With (32), a bound Coulomb state, obeying virial $\frac{1}{2}e^2/r_0$, will absorb a frequency $\lambda_0$, deriving from

$$\tfrac{1}{2}\lambda_0/r_0=4\pi/\alpha=2\cdot 861=1722 \approx 1836 \qquad (33)$$

a field scale factor, perfectly compatible with inverse recoil 1836/1.

---

[21] Instead of R in cm$^{-1}$, recoil for $1/R$ in cm or Å, equal to $10^8/109677{,}58=911{,}7633$ Å, gives $911{,}7633/1836{,}15267 = 0{,}49656$ Å of the expected numerical magnitude, reminding $r_H=r_e+r_p=r_e(1+m_e/m_p)$ Å, with $R=\tfrac{1}{2}e^2/r_e$ cm$^{-1}$ but $1/R=2r_e/e^2$ cm. We return to this problem elsewhere [26].

These strange but unavoidable consequences for recoil, fine structure constant and number π must all be seen in a context of H polarization and of the de Broglie standing wave equation (1e). Combining these intriguing elements and using first principles will result in a generic, system independent polarization theory, easily applied to H and compatible with the equations above [26].

**Conclusion**

For more than 50 years and without any difficulty, theorists admitted that the H Lyman $np_{1/2}$-series is based on numbers (ratios, proportions, symmetries) 1,5 and 3 deriving from Sommerfeld-Dirac QED equation (10). If so, theorists should have no difficulty either to admit that a slightly different H Lyman $ns_{1/2}$-series relies on slightly different numbers ½π (1,57) and π (3,14), given already away with de Broglie's equation, when interpreted in terms of polarization angles (phases). These small differences account for the observed Lamb-shifts and were, eventually, responsible for major new developments in theoretical physics, where polarization remains a central issue. There is a world of difference between the two sets of critical data for the same stable system H: only with de Broglie or Lamb-numbers ½π and π, natural phenomenon polarization appears for the H Coulomb bond but, for reasons extremely difficult to understand, this was never appreciated in the past [1-5]. In essence, Sommerfeld and Kratzer were the pioneers of SUSY (super-symmetry, connecting fermion and boson symmetries), whereas Dirac used electron spin, not even discovered before 1920, and Heisenberg-Schrödinger quantum mechanics, presented in the 1920s. It is still difficult today to understand[22] why Sommerfeld found this strange solution for the bound electron in H.

On the experimental side, H polarization theory must have access to as much as possible precisely measured H Lyman terms, more than the few available today. For instance, the very precise line for H interval 1S-2S [37] is simply a natural H-line, since n=2<π.

Since line profiles are not only affected by PDW shifts, line intensities should be measured also with great precision, since it proves extremely difficult to calculate intensities with present bound state QED. If there is any logic in the analysis above, H polarization must affect H line intensities, which should also exhibit critical behavior (fluctuations) at the critical n-values, given above. In the end, a family of related H lines with the same rotational symmetry is more useful to disclose H chiral behavior than a single line[23]. This puts question marks on the artificial H-experiments at CERN to get at single line 1S-2S, as argued before [5]. Rather than pursuing their impossible dream with artificial H, the physics community would be better served with many more H lines and their profiles, measured with the greatest precision possible.

---

[22] In 1983, Biedenharn wrote that Sommerfeld's derivation of the correct energy of the bound electron in H, 'rediscovered exactly by Dirac', could only be a lucky accident, a sort of cosmic joke at the expense of serious minded physicists, since Sommerfeld's method had no place at all for electron spin…[38]. This is a clear denigration of the merit and validity of the Sommerfeld-Kratzer approach. We proved that this super-symmetric approach uses fermion with boson symmetries instead of electron spin. We stress that this conventional attitude of most theoretical physicists like Biedenharn does great injustice to Sommerfeld, in line with the remarks in footnotes 12, 13, 17 and 19.

[23] Trying to determine handedness for a single line requires that absolute handedness is assessable, which is still impossible.

The history of quantum theory before Schrödinger's time (Bohr-Sommerfeld-Kratzer-Compton-de Broglie) and the origin of bound state QED, the observation of the Lamb shifts, the CBR line,… easily falsify the unjust fate of natural H. Although QFT prescribes H Mexican hat curves and uses these typical bound state patterns at length in the SM to understand elementary particle behavior and chirality, it has persistently failed to make this obvious connection for the most essential [16] element hydrogen: H quartics, H polarization and H chirality must all be intimately and quantitatively connected.

It is undeniable that errors of 2D Bohr $1/n^2$ fermion theory classify as sinusoidal polarization dependent wavelength (PDW) shifts, in line with Compton's non-resonant X-ray experiments and de Broglie's equation. Unambiguous spectral signatures for natural H, ½π and π, only show for the Lamb-shifted H Lyman $ns_{1/2}$ series. While QED uses the very same quartic-generating double square root Sommerfeld-Kratzer function for atom hydrogen, it nevertheless puts a ban on natural H. This widely accepted ban not only contradicts cosmological signatures for dark matter but also classifies QFT as internally inconsistent on neutral antimatter[24].

**Acknowledgements**

**Appendix. Kratzer potential and reduced mass fluctuations in fermion-boson system H**

Chemists know for long that, in composite quantum systems, rotations interfere with vibrations and hope to find a universal potential to explain the order observed for molecular spectroscopic constants [19,20]. In this respect, simple molecular Kratzer potential (6) is superior to Morse's [19]. Numerically equivalent forms are

$$V(n) = -A/n + B/n^2 = A(-1/n + (B/A)/n^2) = (-A/n)(1-b/n) \quad (A1)$$

Eqn. (A1) appears for running Rydbergs (17) in Sommerfeld H theory (see text). With $A=-e^2$; $n=r$ and $A=hc$; $n=\lambda$, (A1) can apply for Coulomb and radiative fields, either for unit mass systems (m=1) or for systems without specifying particle masses. In this more general view, Coulomb field $-e^2/r$ becomes

$$V(r) = -e^2/r + b/r^2 \quad (A2)$$

Particle mass does not show but must behave in an harmonic way, the problem for this Appendix.

Putting the first derivative of (A1) equal to 0 at critical separation $r_0$ gives $e^2/r^2_0 - 2b'/r^3_0 = 0$, which means

$$b = \tfrac{1}{2}e^2 r_0 \text{ and } V(r_0) = -\tfrac{1}{2}e^2/r_0 \quad (A3)$$

Shifting the asymptote to make the minimum coincide with 0, natural difference potentials for Coulomb $-1/r$ and radiative $+1/\lambda$ fields become

$$V(r) - V(r_0) = -e^2/r + \tfrac{1}{2}e^2 r_0/r^2 + \tfrac{1}{2}e^2/r_0 = (\tfrac{1}{2}e^2/r_0)(1-r_0/r)^2 = A(1-n_0/n)^2$$

$$V(\lambda_0) - V(\lambda) = hc/\lambda - \tfrac{1}{2}hc\lambda_0/\lambda^2 - \tfrac{1}{2}hc/\lambda_0 = (-\tfrac{1}{2}hc/\lambda_0)(1-\lambda_0/\lambda)^2 = -A'(1-n_0/n)^2$$

---

[24] If so, my claim for the discovery of natural H [4] should be validated.

$$-[V(\lambda_0)-V(\lambda)] = A'(1-n_0/n)^2 \tag{A4}$$

the basis behind (6) and which leads to the different parabolic equations, discussed in the text.

Unconstrained by mass, second order Kratzer term $|b/n^2|$ represents in a generic way the harmonic relations between frequency and wavelength $\nu\lambda=c$ for radiation, between frequency and separation $\omega r=v$ for oscillations in linear and between angular frequency $\omega$ and radius $r$ for rotations in circular systems

$$d(-e^2/r)/dr=+e^2/r^2;\ d\nu/d\lambda=-c/\lambda^2;\ d\omega/dr=-v/r^2 \tag{A5}$$

These harmony relations (A5) justify Kratzer's second order term $\pm b/n^2$.

The 2nd derivative for $r=r_0$ gives force constants $k$ for harmonic systems, still without specifying masses. In fact, $d^2V(r)/dr^2=-2e^2/r^3+6b/r^4=-2e^2/r^3+3e^2r_0/r^4$ returns $k$ equal to

$$k_C=+e^2/r^3_0 \text{ and } k_R=+hc/\lambda^3_0 \tag{A6}$$

for attractive harmonic Coulomb and radiative fields, consistent with (A4). With (A6), Kratzer's procedure obeys Hooke's law $-½kr^2$ for harmonic equilibrium, since $-½kr^2_0=-½e^2/r_0$ as it should.

This brings us to Kratzer's treatment of harmonic particle mass. For bound, stable systems with masses in harmonic motion and with periodicities governed by sine and cosine, force constant $k$ for inverse harmonic fields relates to reduced mass $\mu$ with frequency $\omega$

$$k=\mu\omega^2 \text{ or } \omega=\sqrt(k/\mu) \tag{A7}$$

Reduced mass $\mu=m_1m_2/(m_1+m_2)=m_1m_2/M$ applies for any complementary system of **size $R=r_1+r_2$ instead** of $M=m_1+m_2$. The only constraint on particle mass in harmonic systems is that it requires reduced, not total mass. Theoretically possible generalized algebraic reduced mass fields are always of form[25]

$$\mu=Mf(\gamma_\pm) \text{ or } \mu/M=f(\gamma_\pm)=\pm\gamma-\gamma^2 \tag{A8}$$

Upon a real or virtual division of total mass $m_H$, reduced mass is no longer a scalar (see text). As with numerical polarization fields (1a)-(1c), the hidden numerical mass field

$$f(\gamma_\pm)=\pm\gamma-\gamma^2 \tag{A9}$$

automatically imposes parabolic, if not sinusoidal behavior for any complementary mass (quantum) system. As a result, generic parabolic behavior of reduced mass or density fluctuations in (A9) resemble the curve in Fig. 2a for the running H Rydbergs. While $\gamma=0$ and $\gamma=1$ both stand for undivided total mass, but not for a system with zero total mass, $\gamma=½$ is the value for a perfectly symmetrically divided system with mass parts $½M$ each. With fermion-boson symmetries in mind, $\gamma$ small ($\approx 1/1836$) but not equal to zero implies $1-\gamma\approx+1$ for its complementary part, which corresponds with fermion-behavior (Bohr-Kratzer view, rotations). Intermediary $\gamma$, near $\pm½$, gives boson behavior (Kratzer view, oscillations). Looking at atom H as a boson gives reduced mass $\mu_H=¼m_H$. While force constant $k$ is usually considered a constant (like R in Bohr $R/n^2$ theory), this overlooks parabolic reduced mass effects, if not density fluctuations (A8). Harmonic mass-related effects, when real, will be reflected in observable frequencies of harmonic systems, by virtue of (A7).

---

[25] Reduced mass $\mu$ requires harmony for $m_1$ and $m_2$: $\mu=m_1m_2/(m_1+m_2)=m_1m_2/M$ is equivalent with $\mu/M=m_1m_2/M^2$ or $\mu M=m_1m_2$. Scaling parts gives $\mu=(\gamma-\gamma^2)M=\gamma(1-\gamma)M$ since $m_1m_2/(m_1+m_2)^2=m_1(M-m_1/)/M^2=(m_1/M)(1-m_1/M)=(\gamma-\gamma^2)=\gamma(1-\gamma)$ is exact. With the law of association, permutation $+1=-m_1/M+(1+m_1/M)$ is valid but gives $-\gamma(1+\gamma)$ instead. This leads to algebraic reduced masses, as in reduced mass field $f(\gamma_\pm)$ in (A8). An inverse view on mass $m_X=C/r_X$, used in classical and quantum physics, transforms linear mass relation $M=m_1+m_2$ into harmonic size $r_X=r_1r_2/(r_1+r_2)=r_1r_2/r_X$ or $r_X=\sqrt(r_1r_2)$, the basis of (1b) and (1c). The common sense behind complementary mass is that we can decide at will whether or not to divide a system: $M=m_1+m_2$ with $m_1=0$ (no bisection) returns M undivided but makes reduced mass zero (see [X] for a preliminary discussion). We deal with this important case of complementary systems elsewhere [26].

To simplify the discussion and to avoid this extra degree of freedom of parabolic form for reduced, harmonic mass, associated with (A9), we proceed with Kratzer's potential in the hypothesis of constant reduced mass. With $v^2=\omega^2r^2$ and $k_C$ in (A6), equilibrium for an harmonic Coulomb system with constant reduced mass obeys

$$\mu\omega^2r^2_0=\mu v^2=e^2/r_0 \qquad (A10)$$

Therefore, Kratzer's potential (6) not only gives equilibrium condition (A10) for mass-systems when in oscillatory harmonic motion but does not put constraints on the mass-distribution (reduced mass), which can apply for any value of $\gamma$ in the domain $0<\gamma<1$. This includes extreme fermion as well as intermediary boson behavior for atomic system H.

In the former case, (A10) is formally degenerate with Bohr's first equilibrium condition for the rotating electron in the Coulomb field $\mu v^2r_0=e^2$ for radial velocity v on a Coulomb field axis. Bohr's central field approximation

$$V_{Bohr}(r)=-e^2/r +\tfrac{1}{2}\mu v^2=-e^2/r +\tfrac{1}{2}\mu\omega^2r^2 \qquad (A11)$$

led to valid rotator levels $E_n=(-\tfrac{1}{2}e^2/r_0)/n^2=-R/n^2$, constrained by fermion behavior as well as by his quantum hypothesis (29a) for angular frequencies $\omega$ and angular velocities, perpendicular to the field axis. Bohr's potential (A11) does not relate to oscillations between complementary parts along a field axis, applying for boson-type harmonic H. The Kratzer potential is essentially a harmonic field for any harmonic system, divided in any harmonic way.

Kratzer boson-type oscillations in H rely on vibrations between 2 neutral, complementary parts of atom H, with maximum mass $\tfrac{1}{2}m_H$ each. Bisecting H atom mass in the boson way is more classical, since it proceeds on a gravitational basis, difficult to understand at first sight but perfectly in line with classical equilibrium conditions $m_er_e=m_pr_p$ or $m_e/m_p=r_p/r_e$ (see text). In a Coulomb view, bisection is achieved with 2 unit charges on 2 fermions, an extreme bisection of atom H in 2 complementary parts. A Bohr H-state (-,+) with a charge conjugated fermion particle pair $m_e$, $m_p$ (rotations) can nevertheless go over in an intermediate state with neutral boson particle pair $\tfrac{1}{2}m_H$, $\tfrac{1}{2}m_H$ (oscillations), the classical rationale behind Sommerfeld's oscillatory corrections to the orthogonal STR field (see text) to end finally in a charge-inverted fermion state H (+,-), the so-called forbidden state in QFT.

Whatever the meaning of (A9), harmonic Kratzer oscillations of boson type (6) on a single field axis can, nevertheless, interfere with Bohr rotations of fermion type with 2 orthogonal axes. When properly combined and scaled in composite fermion-boson system H, these lead to a Poincaré 3D polarization sphere for H and eventually, to the de Broglie polarization angles (1g), available for almost a century. We return to some of these problems elsewhere [26].

Table 1. *Data for ns and np (terms [14] and energies [15]). Energies used for running Rydbergs $R_H(n)$ in (4c) and for quantum number differences for ns (5a) $10^6\Delta_n$ and (5b) $10^6\Delta'_n$ (all data in cm$^{-1}$)*

| n | Terms[a] ns | Terms np | $-E_{ns}$[b] | $-E_{np}$ | $R_H(n)$ ns | $R_H(n)$ np | $10^6\Delta_n$ ns[c] | $10^6\Delta'_n$ ns |
|---|---|---|---|---|---|---|---|---|
| 1 |  | (0) | (109678,773704) |  | 109678,773704 |  | 0,0000 | 0,0000 |
| 2 | 82258,9559 | 82258,9206 | 27419,8178352 | 27419,8531233 | 109679,271341 | 109679,412493 | 1,1343 | 1,1343 |
| 3 | 97492,2235 | 97492,2130 | 12186,5502372 | 12186,5607410 | 109678,952135 | 109679,046669 | 0,2711 | 0,1808 |
| 4 | 102823,8549 | 102823,8505 | 6854,9188454 | 6854,9232846 | 109678,701526 | 109678,772554 | -0,0823 | -0,0411 |
| 5 | 105291,6329 | 105291,6306 | 4387,1408809 | 4387,1431560 | 109678,522023 | 109678,578899 | -0,2295 | -0,0918 |
| 6 | 106632,1518 | 106632,1505 | 3046,6219504 | 3046,6232675 | 109678,390214 | 109678,437630 | -0,2914 | -0,0971 |
| 7 | 107440,4413 | 107440,4405 | 2238,3324513 | 2238,3332810 | 109678,290114 | 109678,330769 | -0,3149 | -0,0900 |
| 8 | 107965,0517 | 107965,0511 | 1713,7220592 | 1713,7226151 | 109678,211786 | 109678,247368 | -0,3202 | -0,0801 |
| 9 | 108324,7225 | 108324,7221 | 1354,0512214 | 1354,0516120 | 109678,148936 | 109678,180570 | -0,3165 | -0,0703 |
| 10 | 108581,9928 | 108581,9925 | 1096,7809744 | 1096,7812592 | 109678,097442 | 109678,125916 | -0,3083 | -0,0617 |
| 11 | 108772,3435 | 108772,3433 | 906,4302025 | 906,4304165 | 109678,054506 | 109678,080394 | -0,2981 | -0,0542 |
| 12 | 108917,1208 | 108917,1207 | 761,6529040 | 761,6530688 | 109678,018175 | 109678,041907 | -0,2870 | -0,0478 |
| 13 | 109029,7916 | 109029,7914 | 648,9821718 | 648,9823015 | 109677,987041 | 109678,008948 | -0,2759 | -0,0424 |
| 14 | 109119,1923 | 109119,1922 | 559,5814289 | 559,5815327 | 109677,960068 | 109677,980411 | -0,2649 | -0,0378 |
| 15 | 109191,3163 | 109191,3162 | 487,4574955 | 487,4575799 | 109677,936478 | 109677,955466 | -0,2544 | -0,0339 |
| 16 | 109250,3444 | 109250,3443 | 428,4293581 | 428,4294276 | 109677,915674 | 109677,933476 | -0,2445 | -0,0306 |
| 17 | 109299,2655 | 109299,2654 | 379,5082948 | 379,5083528 | 109677,897191 | 109677,913948 | -0,2350 | -0,0277 |
| 18 | 109340,2618 | 109340,2617 | 338,5119774 | 338,5120262 | 109677,880663 | 109677,896489 | -0,2262 | -0,0251 |
| 19 | 109374,9569 | 109374,9569 | 303,8168028 | 303,8168443 | 109677,865795 | 109677,880789 | -0,2178 | -0,0229 |
| 20 | 109404,5791 | 109404,5791 | 274,1946309 | 274,1946665 | 109677,852350 | 109677,866592 | -0,2100 | -0,0210 |

a) series limit 109678,7737 cm$^{-1}$
b) ground state $-E_1$=109678,773704 cm$^{-1}$
c) similar data for np not shown but used for Fig. 2b and Fig. 2c

Table 2. *For series ns$_{1/2}$: theoretical Rydberg fluctuations (26b) H PDW shifts and errors of Bohr theory. Precision test of polarization theory: data with and without fit, see (27b) (all data in cm$^{-1}$)*

| n | $-E_{ns}$[15] | $\Delta R$ (26b)(theo) | H PDW shifts (theo) | Errors[a] Bohr theory | Errors[b] (theo, no fit) | Errors[c] (theo with fit) |
|---|---|---|---|---|---|---|
| 1 | 109678,7737040 | 0,578405 | 0,001069 | (0,00000) | (-0,001065) | (-0,003629) |
| 2 | 27419,8178352 | 0,081759 | 0,124429 | 0,124410 | -0,000018 | -0,000109 |
| 3 | 12186,5502372 | 0,402917 | 0,019617 | 0,019826 | 0,000209 | 0,000010 |
| 4 | 6854,9188454 | 0,654754 | -0,004705 | -0,004511 | 0,000194 | 0,000013 |
| 5 | 4387,1408809 | 0,835058 | -0,010223 | -0,010067 | 0,000156 | 0,000008 |
| 6 | 3046,6219504 | 0,967428 | -0,010777 | -0,010652 | 0,000124 | 0,000005 |
| 7 | 2238,3324513 | 1,067938 | -0,009969 | -0,009869 | 0,000100 | 0,000003 |
| 8 | 1713,7220592 | 1,146580 | -0,008861 | -0,008780 | 0,000081 | 0,000002 |
| 9 | 1354,0512214 | 1,209677 | -0,007780 | -0,007713 | 0,000067 | 0,000001 |
| 10 | 1096,7809744 | 1,261372 | -0,006819 | -0,006763 | 0,000056 | 0,000000 |
| 11 | 906,4302025 | 1,304472 | -0,005992 | -0,005944 | 0,000048 | 0,000000 |
| 12 | 761,6529040 | 1,340942 | -0,005288 | -0,005247 | 0,000041 | 0,000000 |
| 13 | 648,9821718 | 1,372193 | -0,004691 | -0,004655 | 0,000036 | 0,000000 |
| 14 | 559,5814289 | 1,399267 | -0,004183 | -0,004151 | 0,000031 | 0,000000 |
| 15 | 487,4574955 | 1,422945 | -0,003749 | -0,003721 | 0,000028 | 0,000000 |
| 16 | 428,4293581 | 1,443826 | -0,003376 | -0,003352 | 0,000025 | 0,000000 |
| 17 | 379,5082948 | 1,462377 | -0,003055 | -0,003033 | 0,000022 | 0,000000 |
| 18 | 338,5119774 | 1,478966 | -0,002776 | -0,002756 | 0,000020 | 0,000000 |
| 19 | 303,8168028 | 1,493888 | -0,002533 | -0,002515 | 0,000018 | 0,000000 |
| 20 | 274,1946309 | 1,507382 | -0,002320 | -0,002303 | 0,000016 | 0,000000 |

a) Bohr theory, error for 19 levels 379,11 MHz
b) Polarization theory, error for 19 levels without fit 2,04 MHz, improvement by 380/2=190
c) Polarization theory, error for 19 levels with fit 0,24 MHz, improvement by 380/0,25=1520

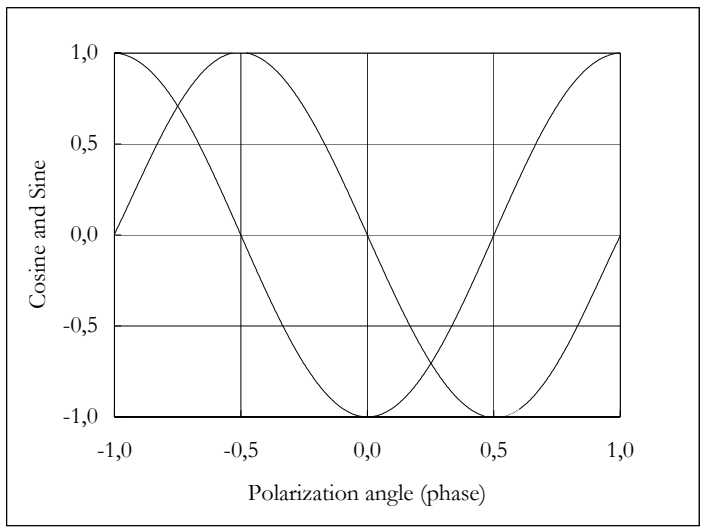

Fig. 1a Generic system independent polarization: cos (full), sin (dashed)

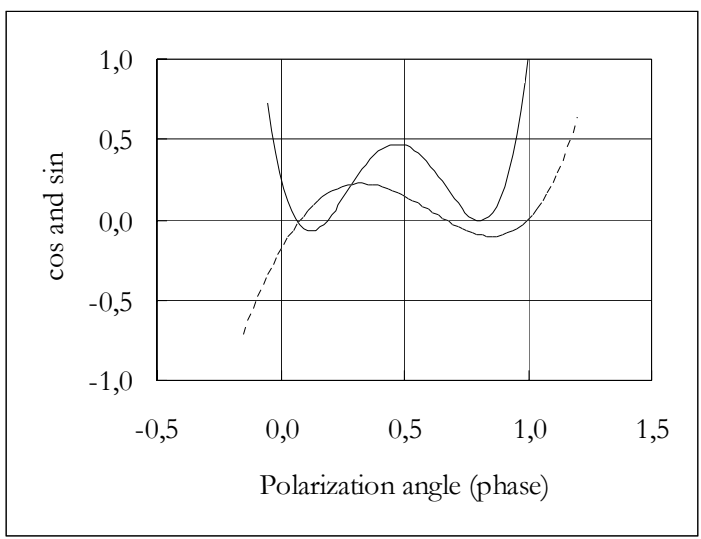

Fig. 1b Generic polarization attenuated by Bohr factor $1/n$: cos (full), sin (dashed)

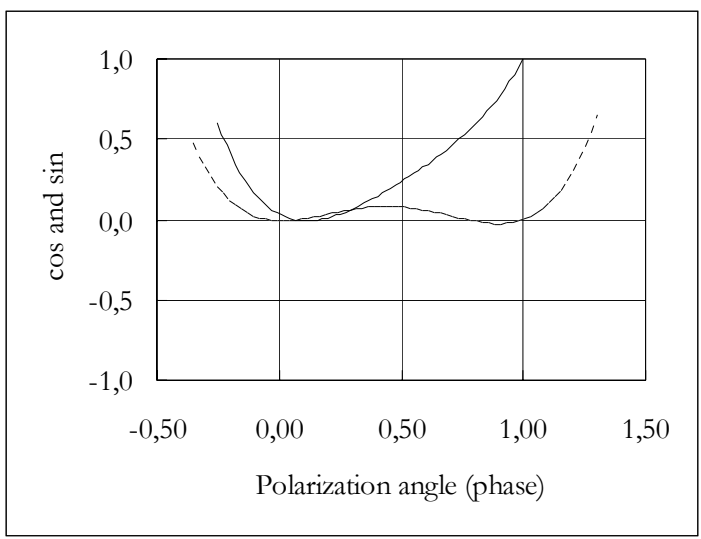

Fig. 1c Generic polarization attenuated by Bohr factor $1/n^2$: cos (full), sin (dashed)

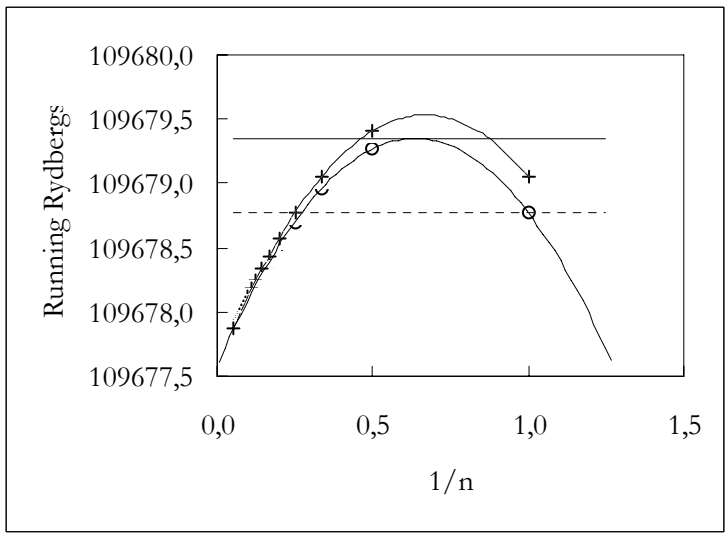

Fig. 2a Running Rydbergs curves: ns (full, extrapolated), np (dashed, not extrapolated), horizontal lines $R_{harm}$ (full), ground state $-E_1=R_1$ (dashed)

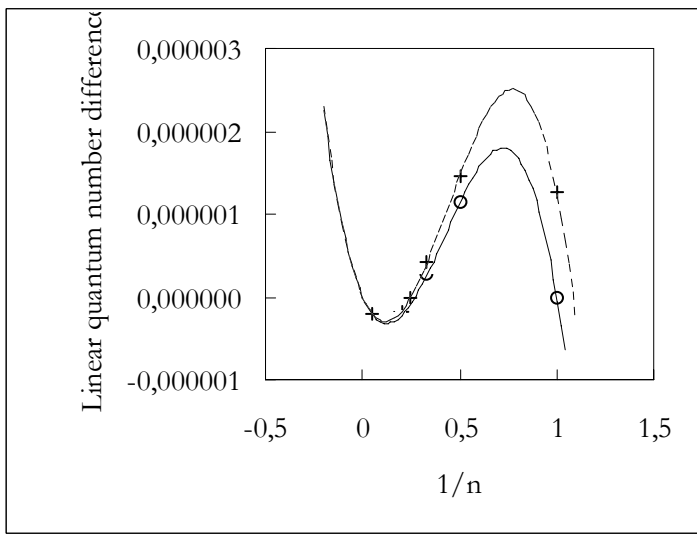

Fig. 2b Linear quantum number differences (5a): ns (full), np (dashed)

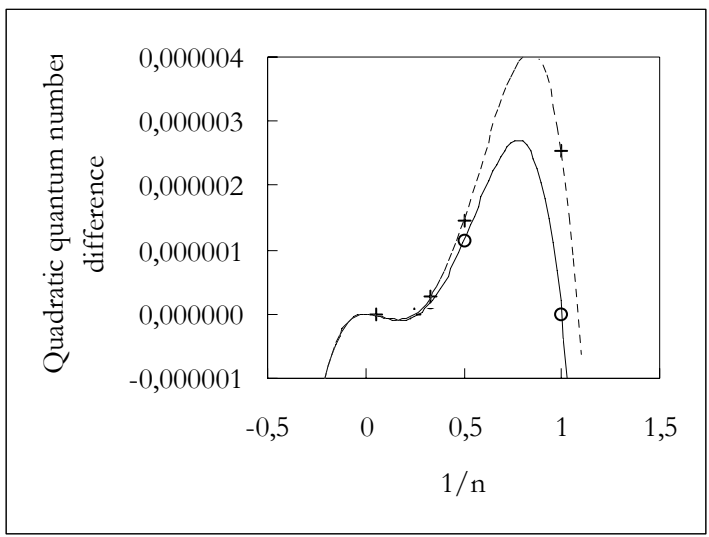

Fig. 2c Quadratic quantum number differences (5b): ns (full), np (dashed)

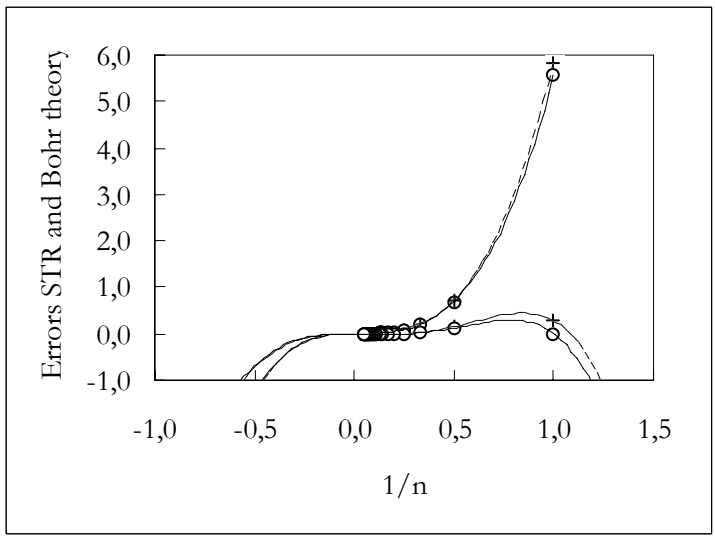

Fig. 2d Large repulsive errors (in cm$^{-1}$) with STR: upper right curves ns (o, full), np (+,dashed) and small sinusoidal errors with Bohr theory: lower right curves ns (o, full) and np (+, dashed)

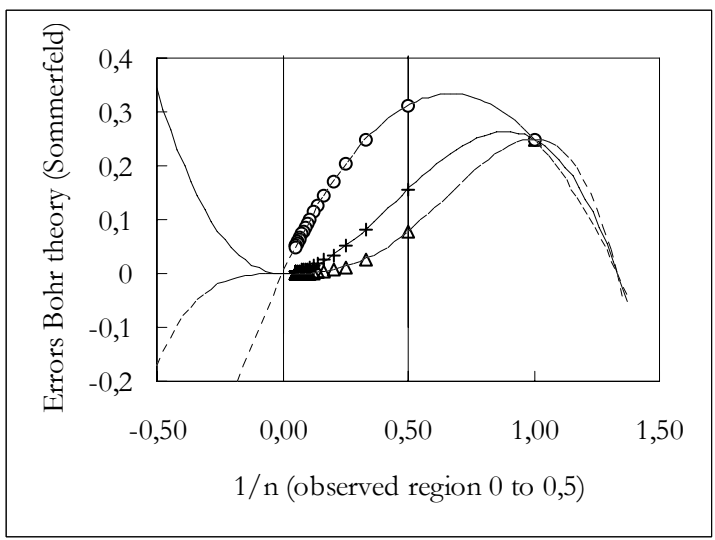

Fig. 3 Extrapolation beyond observed region: from parabola to cubic and to quartic

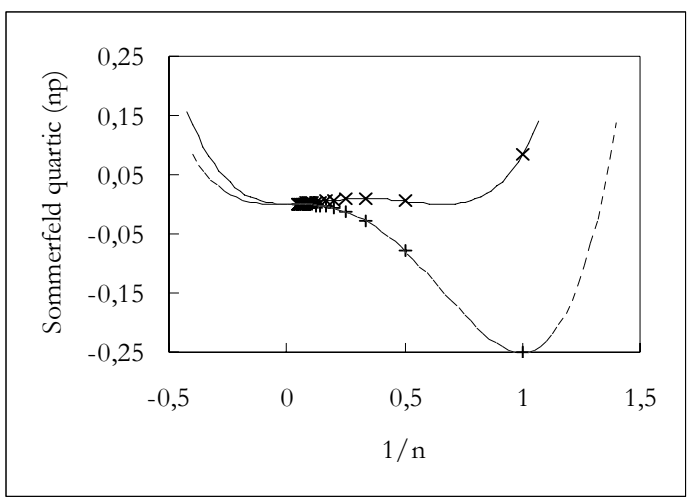

Fig. 4 Effect of asymptote shift on the shape of quartics:
observed with ground state Rydberg (dashed) and theoretical with harmonic Rydberg (full)

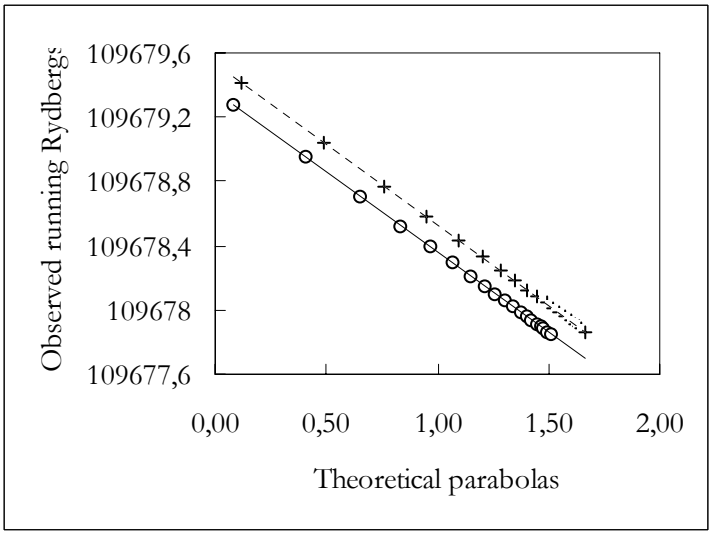

Fig. 5 Observed versus theoretical Rydberg differences:
ns (o, full), fit $R_H(ns)= 109679,353174 - 0,995566 \Delta R_{theo}$
np (+, dashed) fit $R_H(np)= 109679,533968 - 1,001024 \Delta R_{theo}$

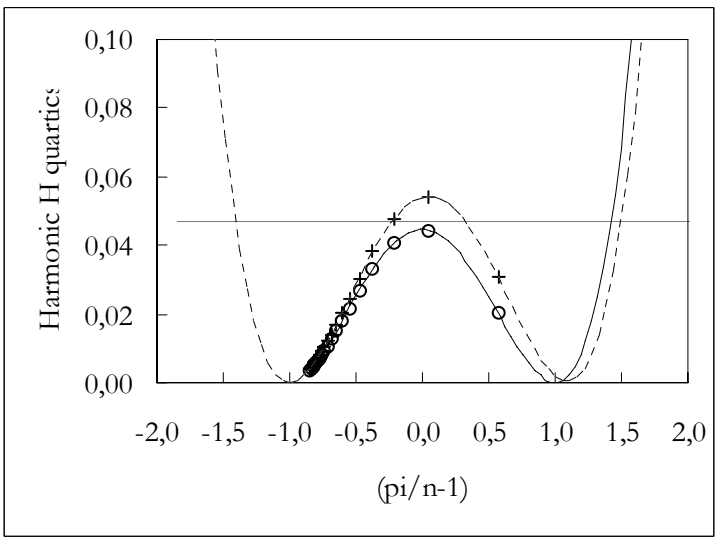

Fig. 6 H PDW shifts, the 21 cm line (horizontal) and harmonic H quartics: ns (o, full), np (+, dashed). Left well: matter domain H, right well: antimatter domain $\underline{H}$.

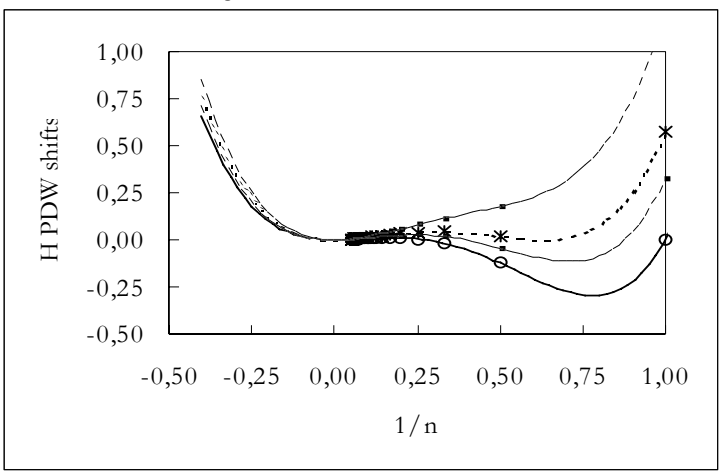

Fig. 7a. Mexican hat curves from HPDW shifts plotted versus 1/n: bottom curve for ground state (o), intermediary (-, dashed), for harmonic R (*, dashed) as in Fig. 6, upper for large R (-, dashed)

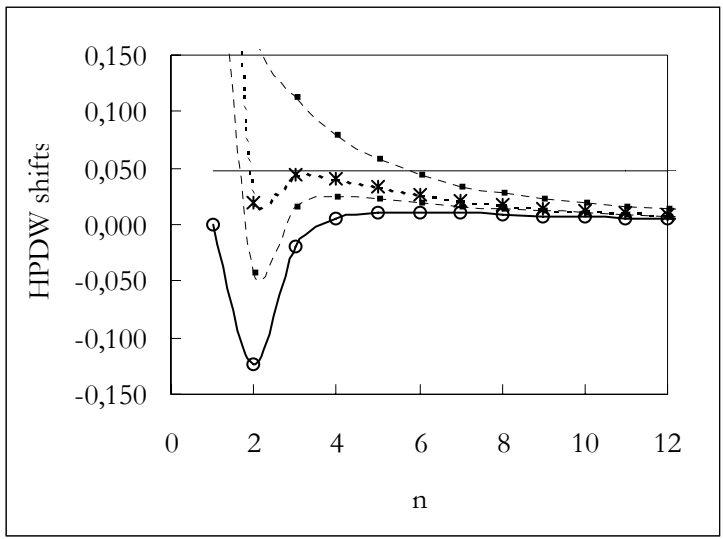

Fig. 7b. Van der Waals-Maxwell curves from HPDW shifts plotted versus n: bottom curve for ground state (o), intermediary (-, dashed), for harmonic R (*,dashed), upper for higher R (-, dashed). Full: 21 cm H line